\documentclass[12pt]{article}
\usepackage[dvips]{graphicx}
\usepackage{psfrag}

\def\lesssim{\mathrel{\mathpalette\vereq<}}
\def\gtrsim{\mathrel{\mathpalette\vereq>}}
\makeatletter
\def\vereq#1#2{\lower3pt\vbox{\baselineskip1.5pt \lineskip1.5pt
\ialign{$\m@th#1\hfill##\hfil$\crcr#2\crcr\sim\crcr}}}
\makeatother

\begin{document}

\begin{titlepage}
\begin{center}
October 19, 2000    \hfill    LBNL-46980 \\
~{} \hfill IASSNS-AST-00/55  \\
~{} \hfill hep-ph/0010231\\

\vskip .1in

{\large \bf On the Evolution of the Neutrino State inside the Sun}

\vskip 0.3in

Alexander Friedland

\vskip 0.05in

{\em Theoretical Physics Group\\
     Ernest Orlando Lawrence Berkeley National Laboratory\\
     University of California, Berkeley, California 94720}

\vskip 0.05in

and

\vskip 0.05in

{\em School of Natural Sciences,
  Institute for Advanced Study\\
  Einstein Drive, Princeton, NJ 08540\footnote{New address from Sept. 1,
2000.}}

\end{center}

\vskip .1in

\begin{abstract}

We reexamine the conventional physical description of the neutrino
evolution inside the Sun.  We point out that the traditional resonance
condition has physical meaning only in the limit of small values of
the neutrino mixing angle, $\theta\ll 1$. For large values of
$\theta$, the resonance condition specifies neither the point of the
maximal violation of adiabaticity in the nonadiabatic case, nor the
point where the flavor conversion occurs at the maximal rate in the
adiabatic case. The corresponding correct conditions, valid for all
values of $\theta$ including $\theta>\pi/4$, are presented. An
adiabaticity condition valid for all values of $\theta$ is also
described. The results of accurate numerical computations of the level
jumping probability in the Sun are presented. These calculations cover
a wide range of $\Delta m^2$, from the vacuum oscillation region to
the region where the standard exponential approximation is good. A
convenient empirical parametrization of these results in terms of
elementary functions is given. The matter effects in the so-called
``quasi-vacuum oscillation regime'' are discussed. Finally, it is
shown how the known analytical results for the exponential, $1/x$, and
linear matter distributions can be simply obtained from the formula
for the hyperbolic tangent profile. An explicit formula for the
jumping probability for the distribution $N_e \propto (\coth(x/l)\pm
1)$ is obtained.

\end{abstract}

\end{titlepage}

\newpage
\setcounter{footnote}{0}
\section{Introduction}


The solar neutrino problem (SNP) is a discrepancy between the measured
values of the solar neutrino flux at different energies
\cite{homestake,GALLEX99,SAGE99,SuperK_daynight,Super-K_spectrum} and
the corresponding predictions of the Standard Solar Model (SSM)
\cite{bp2000}. Not only is the observed flux suppressed, compared to
the SSM predictions, but, if the data from the Homestake experiment
are correct, the degree of suppression varies with energy. The leading
explanation of this phenomenon is that neutrinos have small masses and
the mass and flavor bases in the lepton sector are not aligned, just
like in the quark sector. The resulting neutrino oscillations convert
some of the solar electron neutrinos into another neutrino species.

Neutrino oscillation solutions to the SNP have traditionally been
divided into the so-called Mikheyev-Smirnov-Wolfenstein (MSW)
solutions \cite{W1977,MS1,MS2} and the vacuum oscillation (VO)
solutions, according to the physical mechanism responsible for the
neutrino flavor conversion in each case. In the MSW case the
conversion is caused by neutrino interactions with the solar (and
Earth's) matter, while in the VO case it is due to long-wavelength
neutrino oscillations in vacuum between the Sun and the Earth. Over
time, it has become a tradition to treat the two cases completely
separately, showing results in separate plots (see, for example,
\cite{bksreview,gonzalezgarciajan00}) and using different input
formulas and different codes.

Justifying such a complete separation, however, requires a careful
analysis of the magnitude of the solar matter effects and the degree
of decoherence of vacuum oscillations.  The separation assumption has
been recently reexamined by the author \cite{vacuummsw} and it has
been found that the solar matter effects are nonnegligible for the
vacuum oscillation solutions with $\Delta m^2 \gtrsim 5\times
10^{-10}$ eV$^2$. This conclusion has been subsequently verified by
other authors
\cite{lisiquasivac,Gago2000_3flavor,gonzalezgarciasept00}, and the
term \emph{quasi-vacuum oscillations} (QVO) has been coined to refer
to the region where both effects influence the neutrino survival
probability \cite{lisiquasivac}.

It must be mentioned that the experimental situation has changed since
the QVO solutions were first introduced.  At the time, the most
preferred part of the VO solutions was in the region $\Delta
m^2<10^{-10}$ eV$^2$. The latest Super-Kamiokande spectrum data,
however, disfavors a large fraction of the vacuum oscillation region,
roughly $2\times 10^{-11}$ eV$^2$ $<\Delta m^2 <4\times 10^{-10}$
eV$^2$ \cite{superk_neutrino2000,gonzalezgarciasept00}. At the same
time, the solutions with $\Delta m^2 >4\times 10^{-10}$ eV$^2$,
\emph{i. e.}  the QVO solutions, remain allowed.

Prior to \cite{vacuummsw}, the VO solutions had always been studied in the
range of the neutrino mixing angle $0\leq\theta\leq\pi/4$ for a fixed
sign of $\Delta m^2$.  When matter effects are included, however, this
only covers a half of the full parameter space. To cover the full
space, one can either (i) keep $\theta$ in the range
$0\leq\theta\leq\pi/4$ and consider both signs of $\Delta m^2$, or
(ii) fix the sign of $\Delta m^2$ and vary $\theta$ from $0$ to
$\pi/2$. We advocate the second option as a better \emph{physical}
choice, because it makes manifest the continuity of physics around the
maximal mixing \cite{ourregeneration,ourdarkside}.

The parametrization $0<\theta<\pi/2$ requires one to reexamine the
choice of a variable for plots, because the traditional
variable $\sin^2 2\theta$ is not suitable for this purpose
\cite{ourregeneration}.  While either $\theta$ or $\sin^2\theta$
would be adequate for plotting only the QVO region, neither
choice allows one to take a global view of the neutrino parameter space
and show all solutions, including the SMA and (quasi)vacuum
oscillation solutions, on the same plot \cite{ourdarkside}. A
particularly convenient choice turns out to be $\tan^2 \theta$ on the
logarithmic scale, first used in \cite{FLM1996} to describe 3-family
MSW oscillations. In addition to covering the range $0<\theta<\pi/2$,
it also does not introduce any unphysical singularity around
$\theta=\pi/4$ (unlike the traditional $\sin^2 2\theta$, see
\cite{ourregeneration}) and makes it easy to see where in the vacuum
oscillation region the evolution in the Sun becomes completely
nonadiabatic (points $\theta$ and $\pi/2-\theta$ become equivalent, so
that solutions become symmetric with respect to the $\theta=\pi/4$
line).

In the first part of this paper we address several conceptual
questions that arise in the analysis of the solar matter effects and
become particularly apparent for the values of the mixing angle
$\theta\ge\pi/4$. To introduce these questions, it is useful to first
summarize the basic mechanism of the neutrino evolution in the
Sun. Inside the Sun, because of the changing electron density, the
eigenstates of the instantaneous Hamiltonian change along the neutrino
trajectory. For $\Delta m^2/E_\nu\ll 10^{-5}$ eV$^2$/MeV the neutrino
is produced almost completely in the heavy eigenstate. If the
parameters $\Delta m^2/E_\nu$ and $\theta$ are such that the neutrino
remains in the heavy eigenstate as it travels to the solar surface
(\emph{adiabatic} evolution), there are no subsequent vacuum
oscillations. To oscillate in vacuum, as a necessary condition, the
neutrino must at some radius in the Sun ``jump'' into the
superposition of the heavy ($\nu_2$) and light ($\nu_1$) mass
eigenstates.  Conventional VO regime is realized when this ``jumping''
is \emph{extremely nonadiabatic} (preserving flavor), in which case
the neutrino exits the Sun as
$\cos\theta|\nu_1\rangle+e^{i\phi}\sin\theta|\nu_2\rangle$. In the QVO
regime the neutrino still partially jumps in the $\nu_1$ eigenstate, but
with a smaller amplitude.

The obvious questions one would like to answer are:
\begin{itemize}
\item What physical criteria determine whether
the neutrino evolution is adiabatic or not?
\item At what radius in the Sun does the nonadiabatic ``jumping''
  between the eigenstates of the instantaneous Hamiltonian take place?
\end{itemize}
The traditional wisdom is that one should analyze the density profile
around the so-called resonance point, \emph{i.e.}, the point where the
difference of the eigenvalues of the instantaneous Hamiltonian is
minimal and the local value of the mixing angle is $\theta_m=\pi/4$
(see, \emph{e.g.},
\cite{MS1987jetp,Bethe1986,Haxton1986,Parke1986,KP1989review,
akhmedov1999lectures,haxton2000lectures}).  This, however, clearly
needs to be modified for large mixing angles. In particular, for
$\theta>\pi/4$ the resonance, defined in this way, simply does not
exist.
We will show how this contradiction is resolved in Section
\ref{sect:resonance}. 
In Section \ref{sect:adiabaticity} we formulate the adiabaticity
condition that, unlike the standard result, remains valid for 
$\theta\gtrsim\pi/4$.

In Section \ref{sect:numerical} we present the results of numerical
calculations of the jumping probability $P_c$ for the neutrino
propagating in the realistic solar profile. The calculations are
carried out for a wide range of $\Delta m^2$ and $\tan^2 \theta$, from
the VO region to the region where the exponential density
approximation is valid. We show how the adiabaticity prescription of
Section \ref{sect:adiabaticity} applies to this case. We also give a
simple empirical prescription on how to compute $P_c$ in this range of
the parameters in terms of only elementary functions. Such an
empirical parametrization of the numerical results can be helpful if
one would like to be able to quickly estimate the value of $P_c$
anywhere in the range in question without having to solve the
differential equation each time.

In Section \ref{sect:QVO}, we discuss what happens in the transitional
region between the adiabatic and nonadiabatic regimes (QVO). In
particular, we determine what part of the solar electron density
profile is primarily responsible for the matter effects in this
region.

Finally, in Section \ref{sect:onesolution} we comment on the four
known exact analytical solutions for the neutrino jumping probability
$P_c$. Such solutions have been found for the linear, exponential,
$1/r$, and the hyperbolic tangent matter density profiles. A natural
question to ask is whether these profiles have something in common
that makes finding exact solutions possible. Using the formulation of
the evolution equations introduced in Section \ref{sect:noresonance},
we show that all four results are not independent and that, given the
formula for the hyperbolic tangent profile, one can very simply obtain
the other three solutions. As an added benefit, we obtain an exact
expression for the density distribution $N_e \propto (\coth(x/l)\pm
1)$.

\setcounter{footnote}{0}
\section{Physics of the nonadiabatic neutrino evolution}
\label{sect:noresonance}

\subsection{Review of the oscillation formalism}
\label{sect:intro}


For completeness, we begin by summarizing the well-known basic
formalism for neutrino oscillations in matter. In the simplest case,
when the mixing is between $\nu_e$ and another active neutrino
species, the evolution of the neutrino state is determined by four
parameters: the mass-squared splitting between the neutrino mass
eigenstates $\Delta m^2\equiv m_2^2 - m_1^2$, the neutrino mixing
angle $\theta$, the neutrino energy $E_\nu$, and the electron number
density $N_e$ of the medium. One has to solve the Schr\"odinger
equation $i d\phi/dt=H\phi$, where $\phi=(\phi_e,\phi_\mu)^T$ is the
state vector made up of the electron neutrino and the muon
neutrino.\footnote{In reality, $\phi_\mu$ here denotes a linear
combination of $\phi_\mu$ and $\phi_\tau$ in which $\phi_e$
oscillates.}  The Hamiltonian $H$ is given by \cite{W1977}
\begin{eqnarray}
  \label{eq:basicH}
  H= \mbox{const}+
  \left(\begin{array}{cc}
      A-\Delta \cos 2\theta & \Delta \sin 2\theta \\
      \Delta \sin 2\theta & \Delta \cos 2\theta-A
  \end{array}\right),
\end{eqnarray}
where $\Delta \equiv \Delta m^2/(4 E_\nu)$ and $A \equiv \sqrt{2} G_F
N_e/2$. The constant in the Hamiltonian is irrelevant for the study of
oscillations and will be omitted from now on. The time variable $t$
may be replaced by the distance traveled $x$, since the solar
neutrinos are ultrarelativistic.

For a constant electron number density $N_e$ the Hamiltonian can be
trivially diagonalized, 
$H'=V H V^\dagger=\mbox{diag}(-\Delta_m, +\Delta_m)$, where
\begin{eqnarray}
  \label{eq:Delta_m}
  \Delta_m=\sqrt{(A-\Delta \cos 2\theta)^2+(\Delta \sin 2\theta)^2}=
  \sqrt{A^2-2 A\Delta \cos 2\theta+\Delta^2}.
\end{eqnarray}

In terms of $\Delta_m$, the Hamiltonian (\ref{eq:basicH}) can be
rewritten as
\begin{eqnarray}
  \label{eq:HDelta_m}
  H=
  \left(\begin{array}{rr}
      -\Delta_m \cos 2\theta_m & \Delta_m \sin 2\theta_m \\
      \Delta_m \sin 2\theta_m & \Delta_m \cos 2\theta_m
  \end{array}\right),
\end{eqnarray}
where $\theta_m$ is the mixing angle \emph{in matter}. 
The rotation matrix $V$ is given by
\begin{eqnarray}
  \label{eq:V}
  V=
  \left(\begin{array}{rr}
    \cos\theta_m & -\sin\theta_m \\
    \sin\theta_m & \cos\theta_m 
  \end{array}\right).
\end{eqnarray}

The parameters $\Delta_m$ and $\theta_m$ are related to the original
parameters in the Hamiltonian (\ref{eq:basicH}) as follows
\begin{eqnarray}
  \label{eq:relatesin_m}
  \Delta_m \sin 2\theta_m &=& \Delta\sin 2\theta,\\
  \label{eq:relatecos_m}
  \Delta_m \cos 2\theta_m &=& \Delta\cos 2\theta-A,\\
  \label{eq:relatetan_m}
  \tan 2\theta_m &=& \frac{\Delta\sin 2\theta}{\Delta\cos 2\theta-A}.
\end{eqnarray}

We will always label the light mass eigenstate by $\nu_1$ and the 
heavy one by $\nu_2$. Since one can redefine the phases of
$\nu_{e,\mu}$ and $\nu_{1,2}$, it is easy to see that in this
convention the physical range of the mixing angle is
$0\leq\theta\leq\pi/2$ \cite{ourdarkside,ourregeneration}. 

As long as $N_e(x)$ is constant, the time evolution of the mass eigenstates
is particularly  
simple. Each of the two mass eigenstates evolves only by a phase: 
$|\nu_1(t)\rangle=|\nu_1(0)\rangle \exp(i \Delta_m t)$,
$|\nu_2(t)\rangle=|\nu_2(0)\rangle \exp(-i \Delta_m t)$. If at time
$t=0$ the neutrino state is a linear combination
$a|\nu_1(0)\rangle+b|\nu_2(0)\rangle$, the absolute values of the
coefficients $a$ and $b$ do not change with time, \emph{i.e.}, the
probability for the neutrino to ``jump'' from one Hamiltonian
eigenstate to another, $P_c\equiv |a(t=+\infty)|^2-|a(t=0)|^2$,
is trivially zero.

Consider next the case of a varying electron density. In this case,
in general, one can no longer diagonalize the Hamiltonian in
Eq. (\ref{eq:basicH}). However, one can still speak of the eigenstates
of the instantaneous Hamiltonian (henceforth, ``the matter mass
eigenstates''), and define the jumping probability between those
states.  It turns out that if the electron density changes
sufficiently slowly along the neutrino trajectory (to be quantified
later), the jumping probability vanishes, just like in the constant
density case. This is known as the \emph{adiabatic} evolution.  At the
same time, when the density changes abruptly, the jumping probability
is clearly nonzero.  In particular, if the neutrino crosses a
step--function density discontinuity, the flavor state does not have
any time to evolve, while the mass basis in matter instantaneously
rotates. It is easy to see that in this situation, known as the
\emph{extreme nonadiabatic} evolution, the jumping probability is
given by
\begin{eqnarray}
  \label{eq:extremeNA}
  P_c^{\rm NA}=\sin^2(\theta^{\rm before}-\theta^{\rm after}). 
\end{eqnarray}
In general, for a monotonically varying density $P_c$
lies between 0 and $P_c^{\rm NA}$. 

\begin{figure}[t]
  \begin{center}
    \includegraphics[angle=0, width=1.\textwidth]{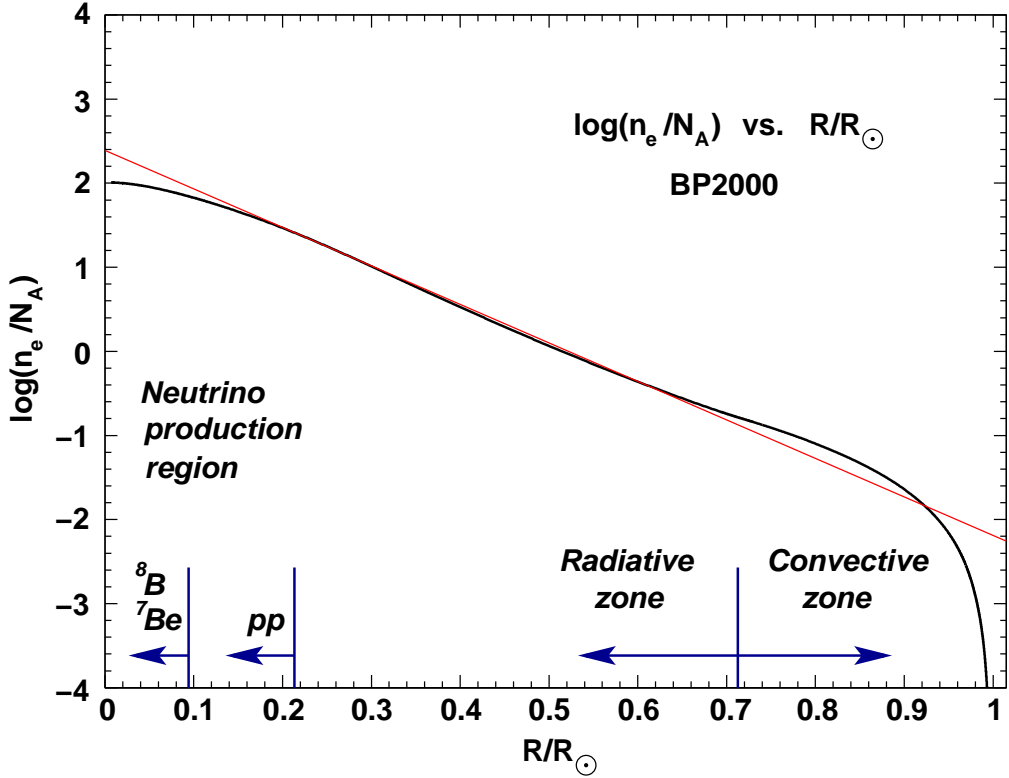}
    \caption{The electron number density profile of the Sun according
      to the BP2000 standard solar model.}
    \label{fig:prfl}
  \end{center}
\end{figure}

In this paper we are concerned with the evolution of solar
neutrinos. The electron number density inside the Sun, $N_e$, falls
off as a function of the distance from the center $r$ as shown in
Fig.~\ref{fig:prfl} \cite{bp2000}. In the range $0.15 R_\odot\lesssim
r \lesssim 0.65 R_\odot$ the profile can be approximated very well by
an exponential $N_e(r)\propto\exp(-r/r_0)$, with
$r_0=R_\odot/10.54=6.60\times 10^4$ km (shown by a straight line in
the Figure). However, in the convective zone of the Sun and also in
the core where the neutrinos are produced the profile deviates rather
significantly from exponential.

In order to study the jumping probability between the matter mass
eigenstates it is convenient to change to the basis these states
define. Substituting in the Schr\"odinger equation $\phi=V^\dagger
\psi$, $\psi\equiv(\psi_1, \psi_2)$, we get
\begin{eqnarray}
  \label{eq:instantH}
  i \frac{d}{d x} (V^\dagger \psi) &=& H V^\dagger
  \psi, \nonumber\\
  i \frac{d\psi}{d x}  &=& V H V^\dagger \psi - 
i V \frac{d V^\dagger}{d x} \psi.
\end{eqnarray}
Since $V H V^\dagger=\mbox{diag}(-\Delta_m,+\Delta_m)$ and from
Eq.~(\ref{eq:V}) 
\begin{displaymath}
  V\frac{d V^\dagger}{d x}=
  \left(\begin{array}{rr}
    0 & 1 \\
    -1 & 0
  \end{array}\right)\frac{d \theta_m}{d x},
\end{displaymath}
we obtain the desired evolution equation in the basis of the matter
mass eigenstates \cite{MS1987jetp}
\begin{eqnarray}
  \label{eq:massbasis}
  \frac{d}{d x}  
  \left(\begin{array}{c}
      \psi_1 \\
      \psi_2
   \end{array}\right) =   
  \left(\begin{array}{cc}
      i \Delta_m & -d\theta_m/d x \\
      d\theta_m/d x & -i \Delta_m
    \end{array}\right) 
  \left(\begin{array}{c}
      \psi_1 \\
      \psi_2
   \end{array}\right).
\end{eqnarray}

The steps outlined so far are standard in the treatment of the MSW
effect. We will next make an extra step that will prove very helpful
for the subsequent analysis, particularly in Section
\ref{sect:onesolution}.  Namely, we will choose $\theta_m$ instead of
$x$ as an independent variable. So long as the density varies
monotonically, such a change is one-to-one. Eq.~(\ref{eq:massbasis})
becomes
\begin{eqnarray}
  \label{eq:thetavariable}
    \frac{d}{d \theta_m}  
  \left(\begin{array}{c}
      \psi_1 \\
      \psi_2
   \end{array}\right) =   
  \left(\begin{array}{cc}
      i \Delta_m/\dot\theta_m & -1 \\
      1 & -i \Delta_m/\dot\theta_m
    \end{array}\right) 
  \left(\begin{array}{c}
      \psi_1 \\
      \psi_2
   \end{array}\right).
\end{eqnarray}
Here $\Delta_m$ and $\dot\theta_m \equiv d\theta_m/d x$ can both be
expressed in terms of $\theta_m$ using the following relationships
\begin{eqnarray}
  \label{eq:thetadot}
  \dot\theta_m&=&\frac{\sin^2 2\theta_m}{2\Delta \sin 2\theta}
  \frac{d A}{d x},\\
  \frac{\Delta_m}{\dot\theta_m} &=& 
  \frac{2\Delta^2 \sin^2 2\theta}{\sin^3 2\theta_m}\frac{1}{d A/d x},\\
  A&=&\frac{\Delta \sin(2\theta_m-2\theta)}{\sin 2\theta_m},  
\end{eqnarray}
which follow directly from
Eqs. (\ref{eq:relatesin_m}-\ref{eq:relatetan_m}). For instance, for
the infinitely extending exponential profile $A(x)=A_0 \exp(-x/r_0)$
the derivative is $d A(x)/d x =-A(x)/r_0$ and so
\begin{eqnarray}
  \label{eq:exponential}
  \frac{\Delta_m}{\dot\theta_m} = 
  -\frac{2\Delta r_0 \sin^2 2\theta}{\sin^2 2\theta_m \sin(2\theta_m-2\theta)}.
\end{eqnarray}

The angle $\theta_m$ varies from its value at the production point 
 $\theta_\odot$ to its vacuum value $\theta$. For the infinite
exponential profile we have $\theta_\odot\rightarrow \pi/2$.
Notice that the quantity $\Delta_m/\dot\theta_m$ in Eq.
(\ref{eq:exponential}) is singular when $\theta_m$ approaches either of
its limiting values, as should be expected. 

\begin{figure}[ht]
  \begin{center}
    \psfrag{theta_m}{{\large $\theta_m$}}
    \psfrag{Dmthetadot}{{\large $|\Delta_m/\dot\theta_m|$}}
    \psfrag{1}{{\large 1}}
    \psfrag{1.5}{{\large 1.5}}
    \psfrag{1.4}{{\large 1.4}}
    \psfrag{1.3}{{\large 1.3}}
    \psfrag{1.2}{{\large 1.2}}
    \psfrag{1.1}{{\large 1.1}}
    \psfrag{0.9}{{\large 0.9}}
    \psfrag{0.8}{{\large 0.8}}
    \psfrag{10}{{\large 10}}
    \psfrag{8}{{\large 8}}
    \psfrag{6}{{\large 6}}
    \psfrag{4}{{\large 4}}
    \psfrag{2}{{\large 2}}
    \psfrag{Exponential}{\textsf{\large Exponential}}
    \psfrag{BP2000}{\textsf{\large BP2000}}
    \includegraphics[angle=0, width=1.\textwidth]{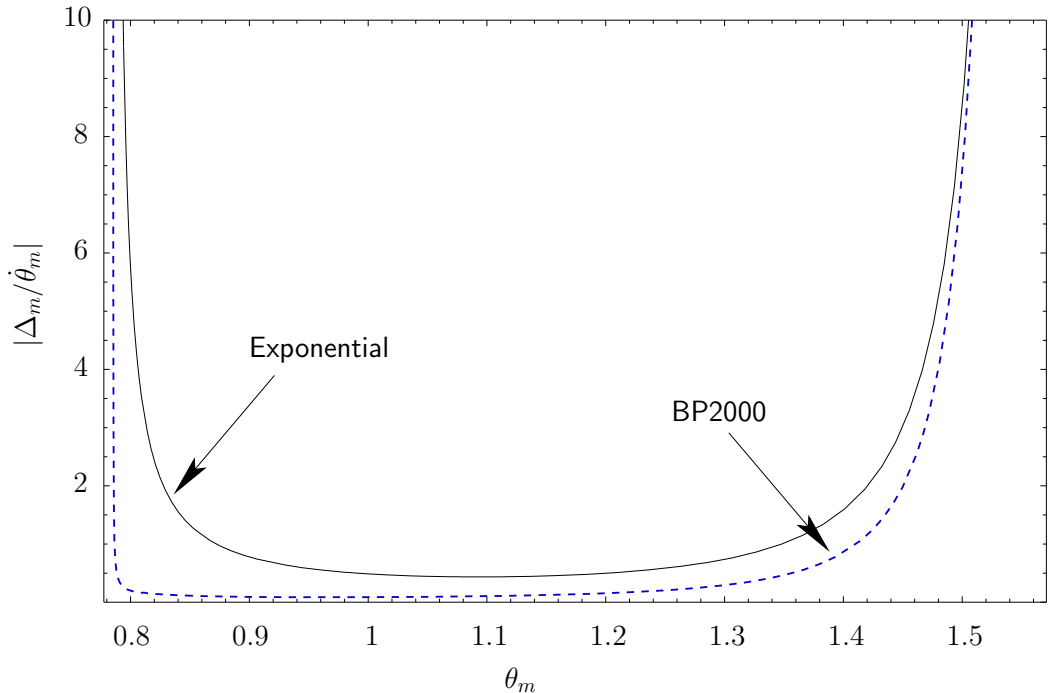}
    \caption{The quantity $|\Delta_m/\dot\theta_m|$ as a function of
      $\theta_m$ for the exponential profile $A(x)\propto
      \exp(-x/r_0)$ (solid line) and BP2000 solar profile (dashed
      line) for $\theta=\pi/4$, $\Delta=10^{-9}$ eV$^2$/MeV.}
    \label{fig:Dmthetadot}
  \end{center}
\end{figure}

The shape of the function $|\Delta_m/\dot\theta_m|$ for $\theta=\pi/4$
and $\Delta=10^{-9}$ eV$^2$/MeV for the idealized exponential profile
is shown in Fig.~\ref{fig:Dmthetadot} by the solid curve.  The value
of $r_0$ was chosen to be $R_\odot/10.54$, the slope of the best fit
line in Fig.~\ref{fig:prfl}. The dashed curve in the Figure shows the
same quantity for the realistic BP2000 solar profile for the same
values of $\theta$ and $\Delta$. It is important to keep in mind that
the two curves change qualitatively differently as one changes
$\Delta$. While the exponential curve just scales by an overall
factor, the BP2000 curve also changes its shape, approaching the shape
of the (rescaled) solid curve for sufficiently large values of
$\Delta$.

\subsection{Modification of the notion of resonance for large~$\theta$.}

\label{sect:resonance}


 
We are ready to address the questions posed in the introduction. 
A convenient starting point is the evolution equation in the form of
Eq. (\ref{eq:thetavariable}). Neutrino is produced in the $\nu_e$
state, which corresponds to $(\cos\theta_\odot,\sin\theta_\odot)$ in
the matter mass basis.  To simplify the presentation, we shall
consider the values of $\Delta\ll 10^{-5}$ eV$^2$/MeV, so that at the
production point $\Delta\ll A_\odot$, and hence
$(\cos\theta_\odot,\sin\theta_\odot)\rightarrow (0,1)$.

The evolution equation can be easily solved in the two limiting
cases. If the off-diagonal elements can be neglected, the evolution is
adiabatic, {\it i.e.}, $|\psi_2(\theta_m)|^2$ remains constant. If, on
the other hand, for almost the entire interval between $\theta_\odot$
and $\theta$ the diagonal terms can be neglected, the solution is
\begin{eqnarray}
  \label{eq:nodiag}
  \psi(\theta) &=& \exp\left[\int_{\theta_\odot}^\theta d\theta_m
    \left(\begin{array}{rr}
      0 & -1 \\
      1 & 0
    \end{array}\right)\right] \psi(\theta_\odot) \nonumber\\
&=&  \left(\begin{array}{rr}
      \cos(\theta-\theta_\odot) & -\sin(\theta-\theta_\odot) \\
      \sin(\theta-\theta_\odot) & \cos(\theta-\theta_\odot)
    \end{array}\right) \psi(\theta_\odot).
\end{eqnarray}
This limit corresponds to the extreme nonadiabatic case. The
corresponding jumping probability equals
$P_c=\sin^2(\theta_\odot-\theta)=\cos^2\theta$, in agreement with
Eq.~(\ref{eq:extremeNA}).

Returning for a moment to the physical $x$--space, we note that no jumping
between the mass eigenstates occurs either in the solar core \cite{W1977} or
in vacuum.  The nonadiabatic evolution takes place in a localized region,
with a center at the point of ``the maximal violation of
adiabaticity''. Our goal next is to establish the location of this
point.

Conventional wisdom says that the adiabaticity condition is
violated maximally at the resonance point
\begin{eqnarray}
  \label{eq:resonance}
  A=\Delta \cos 2\theta,
\end{eqnarray}
where the separation between the energy levels is minimal and
$\theta_m=\pi/4$. This assertion can be found in the early papers
\cite{Bethe1986,Haxton1986,MS1987jetp,MS1987uspekhi}\footnote{A
notable exception is Ref. \cite{Messiah1986}. We do not
agree, however, with the adiabaticity criterion proposed there (see
Section \ref{sect:adiabaticity}).}, as well as in numerous
subsequent reviews on the subject, \emph{e.g.}
\cite{KP1989review,akhmedov1999lectures,haxton2000lectures}. However, in all these papers it is
assumed either explicitly or implicitly that the vacuum mixing angle
$\theta$ is small. It is easy to see that for a large value of the
mixing angle the use of the condition in Eq. (\ref{eq:resonance})
leads to a contradiction.

For small $\theta$, Eq. (\ref{eq:resonance}) is satisfied in a layer
in the Sun where the density is $A(x)\simeq\Delta$. As the value of
$\theta$ increases, Eq. (\ref{eq:resonance}) predicts that the
resonance occurs at lower and lower electron density until, as
$\theta$ approaches $\pi/4$, it moves off to infinity. It is not
obvious how to interpret the last result, as it is physically clear that
no level jumping can occur at infinity where neutrinos undergo
ordinary vacuum oscillations. The difficulty is even more obvious when
$\theta>\pi/4$, in which case the resonance simply never occurs. At
the same time, as already mentioned, in the extreme
nonadiabatic regime the level jumping probability is nonzero for
\emph{any} value of $\theta$ and varies \emph{smoothly} around
$\theta=\pi/4$, $P_c=\cos^2\theta$.

The resolution to this apparent paradox is very simple.  As
Eq. (\ref{eq:thetavariable}) suggests, adiabaticity is maximally
violated at the minimum of $|\Delta_m/\dot\theta_m|$
(\cite{MS1987jetp,MS1987uspekhi}). It is easy to see, however, that
the minimum of $|\Delta_m/\dot\theta_m|$ in general does not reduce to
the condition of Eq. (\ref{eq:resonance}). This can be explicitly seen
on the example of the infinite exponential density
distribution. Differentiating Eq. (\ref{eq:exponential}), one finds 
that the minimum occurs when
\begin{eqnarray}
  \label{eq:minexp1}
  \cot(2\theta_m-2\theta)+2\cot(2\theta_m)=0,
\end{eqnarray}
or 
\begin{eqnarray}
  \label{eq:minexp2}
  A=\Delta \frac{\cos 2\theta+\sqrt{8+\cos^2 2\theta}}{4}.
\end{eqnarray}

Unlike Eq. (\ref{eq:resonance}), Eq. (\ref{eq:minexp2}) states there
is a nonadiabatic part of the neutrino trajectory for all physical
values of $\theta$, including $\theta \ge \pi/4$. While both equations
for small $\theta$ predict that jumping between the local mass
eigenstates occurs around $A=\Delta$, Eq. (\ref{eq:minexp2})
states that for maximal mixing it occurs around
$A=\Delta/\sqrt{2}$, not at infinity, and for $\theta$ close to $\pi/2$
it happens around $A=\Delta/2$, all physically sensible results.

\begin{figure}[t]
  \begin{center}
    \psfrag{psi}{{\large $|\psi_2|^2$}}
    \psfrag{r, 10^6 km}{{\large $x$, $\times 10^6$ km}}
    \psfrag{pi/3}{{\large $\theta=\pi/3$}}
    \psfrag{pi/4}{{\large $\theta=\pi/4$}}
    \psfrag{pi/6}{{\large $\theta=\pi/6$}}
    \psfrag{pi/60}{{\large $\theta=\pi/60$}}
    \psfrag{1}{{\large 1}}
    \psfrag{1.4}{{\large 1.4}}
    \psfrag{1.2}{{\large 1.2}}
    \psfrag{1.0}{{\large 1.0}}
    \psfrag{0.8}{{\large 0.8}}
    \psfrag{0.6}{{\large 0.6}}
    \psfrag{0.4}{{\large 0.4}}
    \psfrag{0.2}{{\large 0.2}}
    \includegraphics[angle=0, width=.945\textwidth]{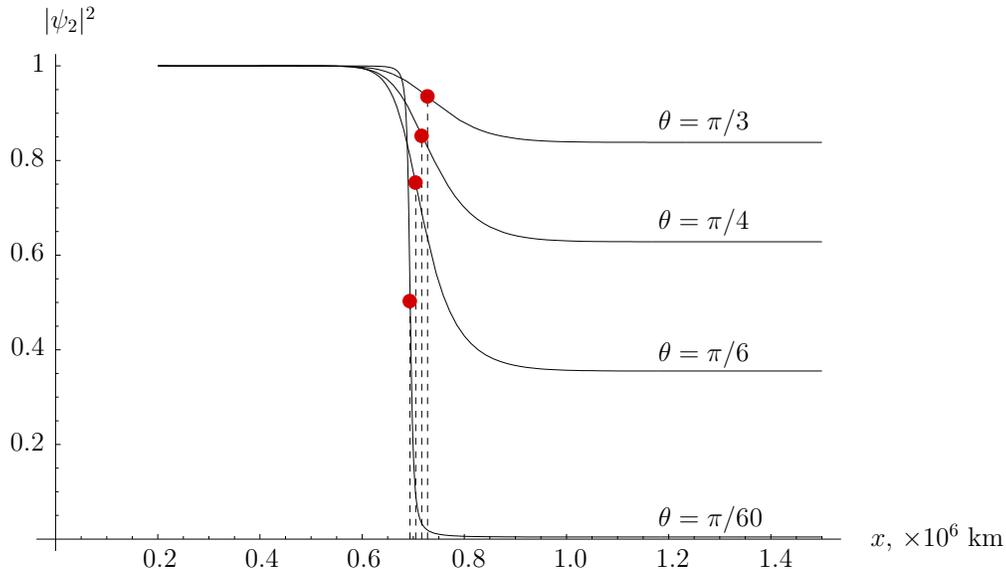}
    \caption{Neutrino state evolution in the case of the 
      infinite exponential density profile for $\Delta m^2/E_\nu=10^{-9}$
      eV$^2$/MeV. The plot shows the probability of finding the
      neutrino in the heavy matter mass eigenstate $\nu_2$ as a function of
      position $x$, for four different values of the vacuum mixing
      angle. The points of the maximal violation of adiabaticity,
      as predicted by Eq. (\ref{eq:minexp2}), are marked.}
    \label{fig:jumpingregion_exp}
  \end{center}
\end{figure}

The situation is illustrated in Fig. \ref{fig:jumpingregion_exp},
which shows the probability of finding the neutrino in the heavy mass
state $\nu_2$ as a function of the distance $x$. The parameters of the
exponential were taken from the fit line in Fig. \ref{fig:prfl} and
$\Delta m^2/E_\nu=10^{-9}$ eV$^2$/MeV. Three large values of the
mixing angle ($\theta=\pi/6,\pi/4,\mbox{ and } \pi/3$) and one small
value ($\theta=\pi/60$) were chosen. The dashed lines and dots mark
the points where adiabaticity is maximally violated, as predicted by
Eq. (\ref{eq:minexp2}). One can see that the partial jumping into the
light mass eigenstate in all four cases indeed occurs around the marked
points.

It is instructive to analyze each of the factors $\Delta_m$ and
$1/\dot\theta_m$ separately. The first one indeed has a minimum at the
traditional resonance point $\theta_m=\pi/4$ (corresponding to
Eq. (\ref{eq:resonance})) or, if $\theta>\pi/4$, at the endpoint
$\theta_m=\theta$. The second one, however, has a minimum at a point
which is, in general, \emph{different} from the resonance. This
minimum exists for all values of $\theta$, including
$\theta>\pi/4$. In the case of the exponential profile, it is located
at
\begin{equation}
  \label{eq:jumpingangle}
  \theta_m=\pi/4+\theta/2,
\end{equation}
or halfway between $\theta$ and $\theta_\odot=\pi/2$. The
corresponding value of the density at that point is 
\begin{equation}
  \label{eq:maxconversion}
  A=\Delta.
\end{equation}
 This
coincides with the resonance condition for $\theta\ll 1$ and that is
why the standard resonance description works very well in this limit. In
general, however, the minimum of the product function lies
somewhere between $\pi/4$ and $\pi/4+\theta/2$ (or $\theta$ and
$\pi/4+\theta/2$, if $\theta>\pi/4$)\footnote{It is even possible for
  certain density profiles and certain values of $\Delta$ and $\theta$
  to have more than one minima.}.

It is also worth mentioning that Eqs. (\ref{eq:jumpingangle}) and
(\ref{eq:maxconversion}) represent an important condition in the
case of the adiabatic evolution. Namely, they specify a point where
the rate of rotation of the mass basis with respect to the flavor
basis is maximal, which in the adiabatic case can be interpreted as a point
where the flavor composition of the neutrino state changes at the
fastest rate. This shows that for large $\theta$
Eq. (\ref{eq:resonance}) not only does not describe the point of the
maximal violation of adiabaticity in the nonadiabatic regime, but also
does not specify the point where the flavor conversion occurs at the
maximal rate in the adiabatic regime.

To summarize, Eq. (\ref{eq:resonance}) can only be used in the small
angle limit. Even in that case one should be careful applying it for
certain purposes. Note, for example, that Eqs. (\ref{eq:resonance})
and (\ref{eq:minexp2}) have different Taylor series expansion around
$\theta=0$,
\begin{eqnarray}
  A &\simeq& \Delta(1-2\theta^2/3)\;\;\;\;\mbox{ for
    Eq. (\ref{eq:minexp2})},
  \nonumber\\
  A &\simeq& \Delta(1-2\theta^2)\;\;\;\;\;\;\;\:\mbox{ for
    Eq. (\ref{eq:resonance})}.
  \nonumber
\end{eqnarray}
Thus, even at small $\theta$, Eq. (\ref{eq:resonance}) fails to predict
how the point of maximal nonadiabaticity \emph{shifts} as a function
of $\theta$.

The belief that jumping between the matter mass eigenstates occurs at
the resonance for all values of $\theta$ might have been one of the
reasons for the tradition to treat separately the cases of
$\theta<\pi/4$ and $\theta>\pi/4$, obscuring the fact that physics is
completely continuous across $\theta=\pi/4$. Over the years, it
has caused some unfortunate confusions, as exemplified by the flawed
criticism of the results of \cite{vacuummsw} in
\cite{NarayanSankar2000}. It was probably the principal reason why the
correct expression for the electron neutrino survival probability in
the $\theta>\pi/4$ part of the QVO region was not given until recently
\cite{ourregeneration,vacuummsw} (\emph{c.f.} Eq. (6) in
\cite{FLM1996}).


One important application of Eq. (\ref{eq:minexp2}) is the
determination of the phase of vacuum oscillations on the Earth
\cite{Petcov88B214,Pantaleone1990}. If jumping
between matter mass eigenstates indeed occurred at the resonance
(\ref{eq:resonance}), one would expect that in the canonical vacuum
oscillation formula
\begin{eqnarray}
  \label{eq:vacsurvival}
    P = 1 - \sin^2 2\theta \sin^2 \left(1.27\frac{\Delta m^2 L}{E}
    +\delta_{\rm res}\right),
\end{eqnarray}
the residual phase $\delta_{\rm res}$ would be minimized when the
distance $L$ was measured from the resonance in the
Sun. Ref. \cite{Pantaleone1990} indeed begins with this assumption,
but after presenting the resulting formulas notes that the residual
phase is much smaller if $L$ is instead measured from the layer where
$A=\Delta$, not $A=\Delta\cos 2\theta$. It is unfortunate that this
important observation has not received proper attention and was not
further developed in the subsequent literature. The preceding
discussion shows that this result is not just a mathematical
coincidence, but has a simple physical explanation.

\begin{figure}[ht]
  \begin{center}
    \psfrag{theta_m}{{\large $\theta_m$}}
    \psfrag{dm/theta_m}{{\large $A(x)/\Delta$ at min of $\Delta_m/\dot\theta_m$ }}
    \psfrag{0}{{\large $0$}}
    \psfrag{1}{{\large $1$}}
    \psfrag{2}{{\large $2$}}
    \psfrag{3}{{\large $3$}}
    \psfrag{0.2}{{\large $0.2$}}
    \psfrag{0.4}{{\large $0.4$}}
    \psfrag{0.6}{{\large $0.6$}}
    \psfrag{0.8}{{\large $0.8$}}
    \psfrag{0.25}{{\large $0.25$}}
    \psfrag{0.5}{{\large $0.5$}}
    \psfrag{0.75}{{\large $0.75$}}
    \psfrag{1.25}{{\large $1.25$}}
    \psfrag{1.5}{{\large $1.5$}}
    \includegraphics[angle=0,width=1.\textwidth]{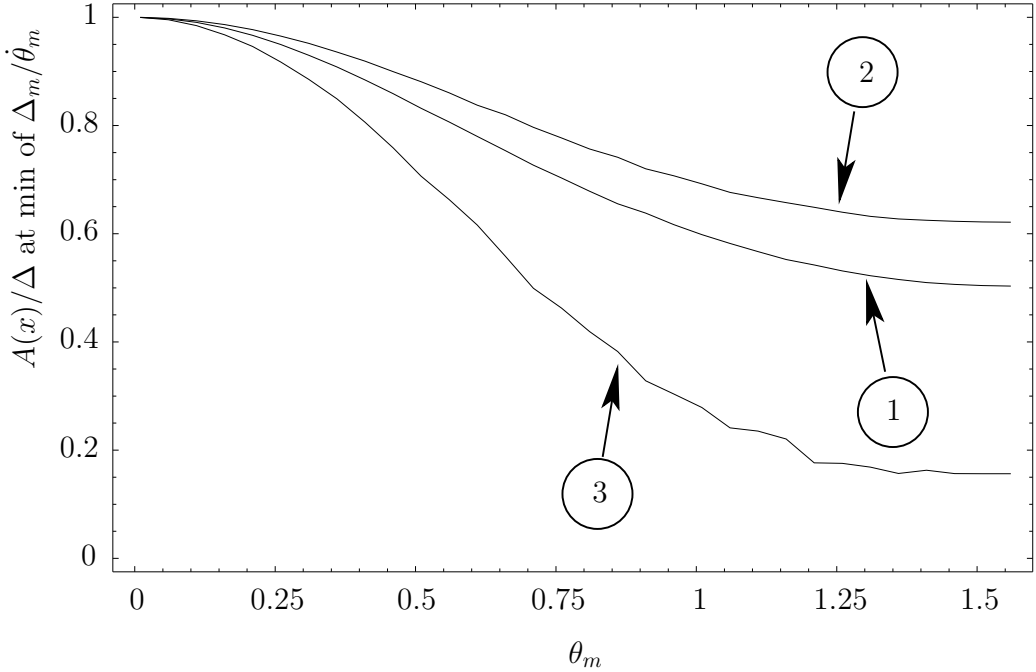}
    \caption{The value of the ratio $A(x)/\Delta$ at the minimum of
      $\Delta_m/\dot\theta_m$ computed for the realistic solar
      density profile (BP2000).  The curves correspond to $\Delta=2\times
      10^{-6}$ eV$^2$/MeV (curve 1), $\Delta=7\times
      10^{-8}$ eV$^2$/MeV (curve 2), and $\Delta=1\times 10^{-8}$
      eV$^2$/MeV (curve 3).}
    \label{fig:bp2000jumping}
  \end{center}
\end{figure}

Finally, it is important to discuss at what density the adiabaticity 
is maximally violated in the case of the realistic solar
profile. Qualitatively, it is not difficult to anticipate the changes
to Eq. (\ref{eq:minexp2}) in this case.
\begin{itemize}
\item For sufficiently large $\Delta$, the adiabaticity is maximally
  violated in the radiative zone, entirely within the exponential part
  of the profile, so that Eq. (\ref{eq:minexp2}) directly
  applies\footnote{As we shall see in Section \ref{sect:numerical}, in
    this range of $\Delta$ the nonadiabatic evolution only occurs
    for $\theta\ll 1$, in which case Eq. (\ref{eq:minexp2}) reduces to
    $A=\Delta$. }.
\item For small $\Delta$, the nonadiabatic part of the trajectory lies
  close to the surface of the Sun where the profile falls off rather
  rapidly. While for small $\theta$ the minimum of
  $\Delta_m/\dot\theta_m$ should still, of course, occur at
  $A=\Delta$, for large angles it is shifted to values of $A$ somewhat
  lower than those predicted by Eq. (\ref{eq:minexp2}). The evolution
  in the latter case will be discussed in more detail in Section
  \ref{sect:QVO}. 
\item For $\Delta$ in the intermediate range, the jumping occurs near
  the bottom of the convective zone where the density falls off somewhat
  slower than in the exponential part. There the value of the ratio
  $A/\Delta$ at large $\theta$ should increase compared to
  Eq. (\ref{eq:minexp2}). 
\end{itemize}
These qualitative expectations are supported by the results of
numerical calculations presented in Fig. \ref{fig:bp2000jumping},
where the ratio $A/\Delta$ at the point of minimal
$\Delta_m/\dot\theta_m$ is plotted as function of $\theta$. The three
curves shown correspond to the values of $\Delta$ in the three
different regimes: $\Delta=2\times 10^{-6}$ eV$^2$/MeV (curve 1),
$\Delta=7\times 10^{-8}$ eV$^2$/MeV (curve 2), and $\Delta=1\times
10^{-8}$ eV$^2$/MeV (curve 3). The numerical studies of the BP2000
solar profile are presented in Section \ref{sect:numerical}.

\subsection{Adiabaticity condition for large $\theta$.}

\label{sect:adiabaticity}


We now turn to formulating the adiabaticity condition that is valid
for all, and not just small, values of the mixing angle $\theta$.  At
first sight it appears that for the evolution to be adiabatic it is
simply enough to require that the diagonal elements in
Eq.(\ref{eq:thetavariable}) be larger than the off-diagonal ones. This
condition becomes most critical at the point where the adiabaticity is
maximally violated.  Since traditionally this point has been
identified with the resonance, a commonly cited condition is
\cite{MS1987jetp}
\begin{equation}
  \label{eq:oldad0}
  |\Delta_m/\dot\theta_m|_{\theta_m=\pi/4}\gg 1.
\end{equation}

Since we have shown that the point of the maximal violation of
adiabaticity in general does not coincide with the resonance,
Eq. (\ref{eq:oldad0}) clearly needs to be modified for large mixing
angles.  Superficially, the problem appears easy to fix: one should
evaluate the left hand side at a value of $\theta_m$ close to the true
point of the maximal violation of adiabaticity. In the case of the
exponential profile, this point would be given by the solution of
Eq. (\ref{eq:minexp1}). For the purpose of an estimate, 
we can approximate it by the point halfway between $\theta$ and
$\pi/2$ (see Eq. (\ref{eq:jumpingangle})),
\begin{equation}
  \label{eq:oldad}
  |\Delta_m/\dot\theta_m|_{\theta_m\sim\pi/4+\theta/2}\gg 1.
\end{equation}

This condition, however, still turns out to be inadequate for large
values of $\theta$. To obtain the correct condition, one has to analyze
the problem more carefully.

The key is to express the information contained in the system of two
evolution equations in a single equation. Such an equation can be easily
written for the ratio $s\equiv\psi_1/\psi_2$. From
Eq. (\ref{eq:thetavariable}) it follows that $s$ obeys the following
first order equation
\begin{equation}
  \label{eq:firstorder}
  \frac{d s}{d \theta_m} = 2 i \frac{\Delta_m}{\dot\theta_m } s - (s^2+1).
\end{equation}

\begin{figure}[htbp]
  \begin{center}
    \psfrag{tan}{{\large $\tan^2 \theta$}}
    \psfrag{dm2}{{\large $\Delta m^2/E_\nu$, (eV$^2$/MeV)}}
    \psfrag{0}{{\large $1$}}
    \psfrag{1}{{\large $10$}}
    \psfrag{2}{{\large $10^2$}}
    \psfrag{3}{{\large $10^3$}}
    \psfrag{-1}{{\large $10^{-1}$}}
    \psfrag{-2}{{\large $10^{-2}$}}
    \psfrag{-3}{{\large $10^{-3}$}}
    \psfrag{7}{{\large $10^{-7}$}}
    \psfrag{8}{{\large $10^{-8}$}}
    \psfrag{9}{{\large $10^{-9}$}}
    \psfrag{10}{{\large $10^{-10}$}}
    \psfrag{11}{{\large $10^{-11}$}}
    \psfrag{Pc=0.9}{{\large $P_c=0.9$}}
    \psfrag{0.7}{{\large $0.7$}}
    \psfrag{0.5}{{\large $0.5$}}
    \psfrag{0.3}{{\large $0.3$}}
    \psfrag{0.1}{{\large $0.1$}}
    \psfrag{0.01}{{\large $10^{-2}$}}
    \psfrag{0.001}{{\large $10^{-3}$}}
    \psfrag{Adiabatic}{\textsf{\Large Adiabatic}}
    \psfrag{Nonadiabatic}{\textsf{\Large Nonadiabatic}}
    \psfrag{Eq1}{{\large Eq. (\ref{eq:oldad0})}}
    \psfrag{Eq2}{{\large Eq. (\ref{eq:oldad})}}
    \psfrag{Eq3}{{\large Eq. (\ref{eq:adiabaticity})}}
    \includegraphics[angle=0,width=1.\textwidth]{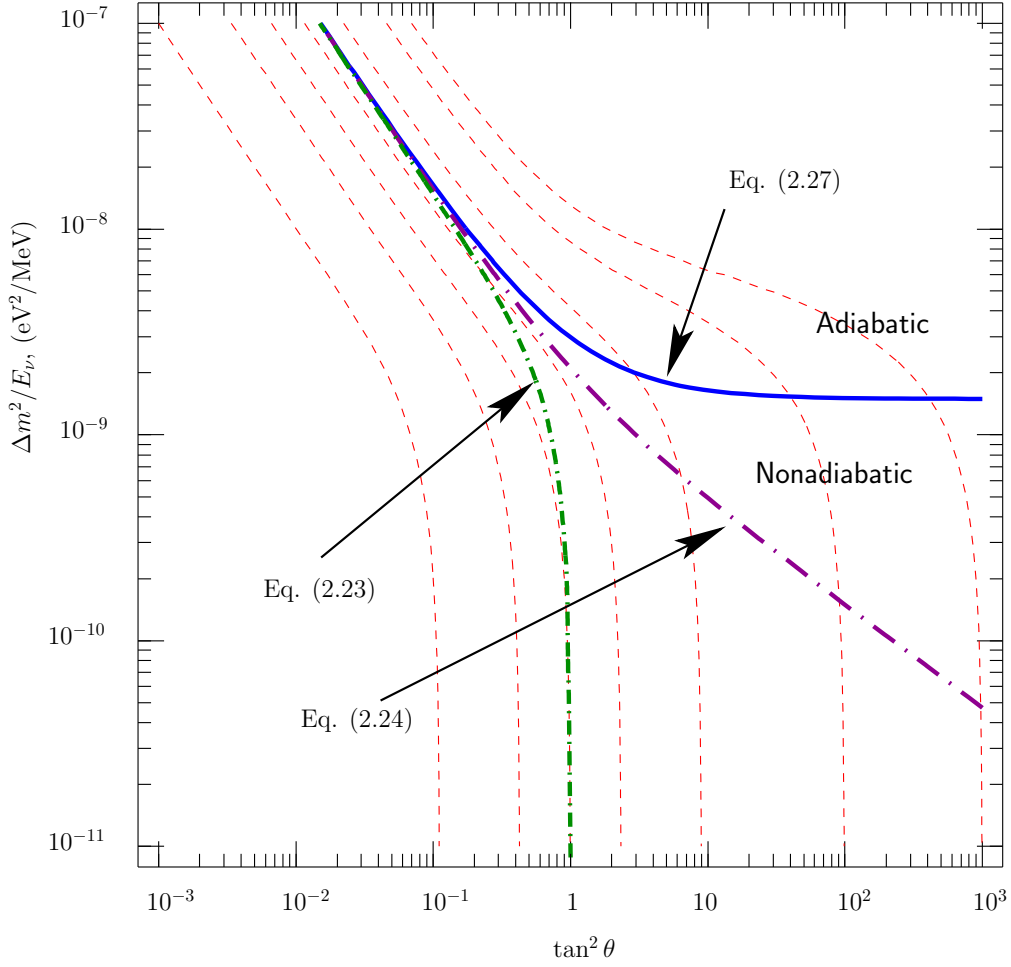}
    \caption{The boundary between the adiabatic and nonadiabatic
      regions in the case of the exponential profile, as predicted by
      Eq. (\ref{eq:oldad0}), Eq. (\ref{eq:oldad}), and
      Eq. (\ref{eq:adiabaticity}). The dashed curves
      show the contours of constant $P_c$. For large angles
      Eq. (\ref{eq:adiabaticity}) provides a better description of
      the boundary.}
    \label{fig:adiabaticity}
  \end{center}
\end{figure}

It is easy to see that by neglecting appropriate terms on the right
hand side one obtains both the adiabatic and nonadiabatic limits. The
adiabatic limit corresponds to neglecting the terms in parentheses,
while the extreme nonadiabatic limit is obtained if one neglects the
first term on the right. In the second case the solution is
$s=\cot(\theta_m)$.  Thus, the self consistent condition to have the
extreme nonadiabatic solution is $2|\Delta_m/\dot\theta_m
s|\ll(s^2+1)$, with $s=\cot(\theta_m)$,
$\theta_m=\pi/4+\theta/2$. In the opposite limit, the evolution is
adiabatic. Thus, we obtain the adiabaticity condition 
\begin{eqnarray}
  \label{eq:adiabaticity0}
  |\Delta_m/\dot\theta_m|_{\theta_m\sim\pi/4+\theta/2} \gg 
(\tan(\pi/4+\theta/2)+\cot(\pi/4+\theta/2))/2,
\end{eqnarray}
or, upon simplification,
\begin{eqnarray}
  \label{eq:adiabaticity}
  \cos\theta |\Delta_m/\dot\theta_m|_{\theta_m\sim\pi/4+\theta/2} \gg 1.
\end{eqnarray}

Notice, that all three conditions, Eqs. (\ref{eq:oldad0}),
(\ref{eq:oldad}), and (\ref{eq:adiabaticity}) agree in the small angle
limit. For large angles, however, especially for
$\theta\gtrsim\pi/4$, they give very different predictions and only
Eq. (\ref{eq:adiabaticity}) is the correct one.

Let us demonstrate this on the example of the infinite exponential
profile. In this case the exact analytical solution is known
\cite{Toshev,petcovanalyt} (see also \cite{kaneko}),
\begin{equation}
  \label{eq:expPc}
  P_c=\frac{e^{\gamma \cos^2\theta}-1}{e^{\gamma}-1}.
\end{equation}
Eq. (\ref{eq:expPc}) has two regimes. For small mixing angles, the
formula reduces to $P_c=\exp(-\gamma \sin^2\theta)$, so the
evolution is nonadiabatic when $\gamma \theta^2=4\pi r_0\Delta
\theta^2 \ll 1$. By contract, for $\theta \gtrsim \pi/4$ and
$\gamma\gg 1$ the numerator is always much smaller than the
denominator. The transition between the adiabatic and nonadiabatic
regimes now occurs when $\gamma \sim 1$, with a weak dependence on
$\theta$.

Let us see if Eq. (\ref{eq:adiabaticity}) correctly captures this
behavior. Using Eq.~(\ref{eq:exponential}) we obtain
\begin{eqnarray}
  \label{eq:nonad_simple}
  8\Delta r_0 \sin^2\theta \gg 1.
\end{eqnarray}
Fig. \ref{fig:adiabaticity} shows the contour of $8\Delta r_0
\sin^2\theta=1$ computed for $r_0=R_\odot/10.54$ (solid line). For
comparison, the dash-dotted curves shows the corresponding prediction
of Eqs. (\ref{eq:oldad0}) and (\ref{eq:oldad}). The dashed curves are
the contours of constant $P_c$ computed using Eq. (\ref{eq:expPc}). It
is clear from the Figure that the description of
Eq. (\ref{eq:adiabaticity}) is correct not only for small $\theta$,
but also for $\theta\gtrsim \pi/4$. The transition from adiabaticity
to nonadiabaticity for $\theta\gtrsim \pi/4$ occurs for $\Delta
m^2/E_\nu \sim 10^{-9}$ eV$^2$/MeV, precisely where $\gamma\sim 1$.

The true usefulness of Eq. (\ref{eq:adiabaticity}) is not so much in
being able to explain the physics behind the known analytical solution
as it is in being able to make predictions for a variety of new
density profiles, provided those profiles are sufficiently
smooth. Such an analysis, however, would be beyond the scope of this
paper.  The question we do want to address is what
Eq. (\ref{eq:adiabaticity}) predicts for the realistic solar density
profile.  It turns out that for large mixing angles and the values of
$\Delta$ that yield nonadiabatic evolution the nonadiabatic jumping
takes place mostly in the convective zone, where the density profile
deviates from the exponential, and so one can no longer rely on
(\ref{eq:expPc}) to get the shape of the boundary between the
adiabatic and nonadiabatic regions. Eq. (\ref{eq:adiabaticity})
nonetheless works quite well. The corresponding numerical
results are presented next.

\setcounter{equation}{0}
\setcounter{footnote}{0}
\section{Calculations for the realistic solar profile}
\label{sect:numerical}

\subsection{Numerical calculations of the jumping probability}
\label{sect:Pc}




In this Section we present the results of numerical computation of the
jumping probability $P_c$ with the realistic (BP2000) solar density
profile.  In practice, ultimately, one would like to compute expected
event rates at various experiments, and for that one needs to know the
solar neutrino survival probability $P(\nu_{e}\rightarrow
\nu_{e})$. It is, however, quite straightforward to show that, for a
given neutrino energy, the survival probability is given by
\begin{eqnarray}
  \label{eq:Pee}
\lefteqn{  P(\nu_{e}\rightarrow \nu_{e}) =
        P_{1} \cos^{2} \theta + (1-P_{1}) \sin^{2} \theta} \nonumber \\
        &-& \sqrt{P_{c} (1-P_{c})} \cos 2\theta_\odot \sin 2 \theta
        \cos \left( 2.54 \frac{\Delta m^{2}}{E}L + \delta \right),
\end{eqnarray}
where $P_1 = P_c \sin^2 \theta_\odot + (1-P_c) \cos^2\theta_\odot$
\cite{ourdarkside}. Thus, the problem of finding $P(\nu_{e}\rightarrow
\nu_{e})$ reduces to finding the jumping probability $P_c$. 

The quantity $P_c$ can be found analytically in several limiting
regimes.  For $\Delta m^2/E_\nu\gtrsim 10^{-7}$ eV$^2$/MeV the
condition of Eq. (\ref{eq:minexp2}) is satisfied well inside the Sun
where the profile is exponential with $r_0=R_\odot/10.54$. Hence,
in this case the jumping probability should
be adequately described by Eq. (\ref{eq:expPc}).

The second regime is $\theta\ll 1$, in which case the standard
resonance description applies and, importantly, the resonance is very
narrow. As a result, the jumping probability can be adequately
described by the analytical formula for a linear density profile,
$P_c= \exp(-\pi \Delta^2\sin^2 2\theta
|dA(x)/dx|_{A=\Delta}^{-1} )$ (Eq. (\ref{eq:linearPc}) of
Sect. \ref{sect:onesolution}).
The contours of constant $P_c$ in this regime are expected to follow
the behavior of the changing slope of the BP2000 density profile shown
in Fig. \ref{fig:prfl}.

The third regime is the region of small $\Delta$ and large
$\theta$. As already discussed, for $\Delta\rightarrow 0$ the
evolution becomes extremely nonadiabatic, {\it i. e.},
$P_c\rightarrow\cos^2\theta$. For $\Delta m^2/E_\nu \sim 10^{-9}$
eV$^2$/MeV we have the so-called quasivacuum oscillation region.  It
was shown in \cite{vacuummsw} that for $\Delta m^2/E_\nu \lesssim
5\times 10^{-9}$ eV$^2$/MeV Eq. (\ref{eq:expPc}) provides a good fit
to numerical calculations, provided one takes $r_0=R_\odot/18.4$. This
region will be discussed in more detail in Section \ref{sect:QVO}.

We next present numerical results for $P_c$ that cover the entire
range between these regimes. We numerically solve
Eq. (\ref{eq:firstorder}) on a grid of points in the range
$10^{-3}<\tan^2\theta<10$, $10^{-11}\mbox{ eV}^2/\mbox{MeV}<\Delta
m^2/E_\nu<2\times 10^{-7}\mbox{ eV}^2/\mbox{MeV}$. Notice that, while
the same result would be obtained by solving the system of equations
in Eq. (\ref{eq:massbasis}), Eq. (\ref{eq:firstorder}) requires
significantly less computer time. Indeed, Eq. (\ref{eq:massbasis})
contains four real functions (real and imaginary parts of $\psi_1$ and
$\psi_2$), but only two of them are independent, since
$|\psi_1|^2+|\psi_2|^2=1$ and the overall phase has no physical
meaning.  

\begin{figure}[htbp]
  \begin{center}
    \psfrag{tan^2 theta}{{\Large $\tan^2 \theta$}}
    \psfrag{dm^2/E}{{\Large $\Delta m^2/E$ (eV$^2$/MeV)}}
    \psfrag{-7}{{\large $10^{-7}$}}
    \psfrag{-8}{{\large $10^{-8}$}}
    \psfrag{-9}{{\large $10^{-9}$}}
    \psfrag{-10}{{\large $10^{-10}$}}
    \psfrag{-11}{{\large $10^{-11}$}}
    \psfrag{1}{{\large $1$}}
    \psfrag{10}{{\large $10$}}
    \psfrag{0.1 }{{\large $0.1$}}
    \psfrag{0.01}{{\large $0.01$}}
    \includegraphics[angle=0,
    width=1.\textwidth]{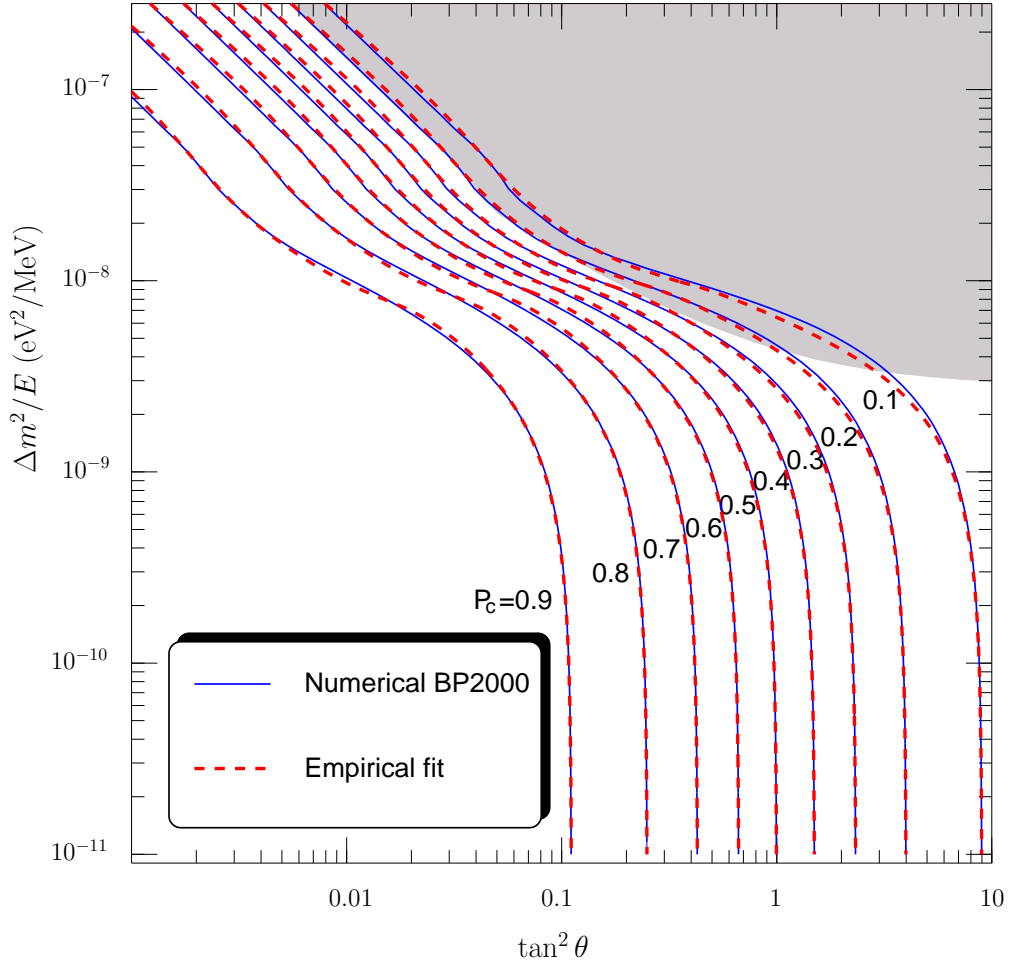}
    \caption{Contours of constant level jumping probability $P_c$. The
      solid line represent the result of numerical calculations using
      the BP2000 solar profile. The dashed lines illustrate a possible
      empirical fit using only elementary functions
      (Eq. (\ref{eq:fit1})). } 
    \label{fig:empiricalfit}
  \end{center}
\end{figure}

The resulting contours of constant $P_c$ are shown in
Fig. \ref{fig:empiricalfit} by solid curves. All the features
anticipated in the discussion above are clearly present.

We can now test the validity of the adiabaticity criterion introduced
in the previous Section (Eq. (\ref{eq:adiabaticity})). The shaded
region in Fig. \ref{fig:prfl} corresponds to 
$\cos\theta |\Delta_m/\dot\theta_m|_{\theta_m=\pi/4+\theta/2} \ge
1$. As one can see by comparing its shape with that of the contours of
constant $P_c$, the criterion in question indeed describes the
adiabatic region quite well, even for $\theta>\pi/4$.

For practical calculations it is convenient to have a simple expression
involving only elementary functions that provides a satisfactory means
of estimating $P_c$ without having to solve the differential equation
each time. We can construct such a purely empirical fit function by
taking Eq. (\ref{eq:expPc}) as a starting point.  Since for both large
and small values of $\Delta m^2/E_\nu$ one can use
Eq. (\ref{eq:expPc}) with the appropriate values of $r_0$, we can
modify Eq. (\ref{eq:expPc}) by making $r_0$ a function of $\Delta
m^2/E_\nu$, smoothly interpolating between $R_\odot/10.54$ and
$R_\odot/18.4$, 
\begin{equation}
  \label{eq:fit1}
  r_0(\Delta m^2/E_\nu) =\frac{R_\odot}{10.54}
\left[0.75 \left(\frac{1}{\pi}\arctan[-10(a+8)]+
\frac{1}{2}\right)+1\right]^{-1}, 
\end{equation}
where $a=\log_{10}[(\Delta m^2/E_\nu)/({\rm eV}^2/{\rm MeV})]$.
This provides an adequate fit for both $\Delta m^2/E_\nu\lesssim 5\times
10^{-9}$ eV$^2$/MeV and $ \Delta m^2/E_\nu \gtrsim 10^{-7}$ eV$^2$/MeV.
By making additional modifications to the function it is possible to
obtain a reasonable fit also to the contours in the intermediate region,
\begin{eqnarray}
  \label{eq:fit2}
  r_0^{\rm fit}(\Delta m^2/E_\nu) &=& 
\frac{R_\odot}{10.54}
\left[0.75 \left(\frac{1}{\pi}\arctan[-10(a+7.95)+0.9b]+
\frac{1}{2}\right)\right. \nonumber\\
&-& \left. 0.3\exp\left(-\frac{(a+7.7)^2}{0.5^2}\right)+1\right]^{-1}, 
\end{eqnarray}
where $b=\log_{10}[\tan^2 \theta]$.  It is important to emphasize that
this equation should only be viewed as a purely empirical fit to the
numerical results, intended to facilitate practical computations of
the solar neutrino survival probability.  

The contours of constant $P_c^{\rm fit}$ computed with
Eq. (\ref{eq:fit2}) are shown by dashed lines in
Fig. \ref{fig:empiricalfit}. The discrepancy between the 
numerical results and $P_c^{\rm fit}$ is $|P_c^{\rm fit}-P_c^{\rm
BP2000}|<0.022$.

\subsection{Matter effects in the QVO region.}
\label{sect:QVO}

\begin{figure}[htbp]
  \begin{center}
    \psfrag{theta_m}{{\large $\theta_m$}}
    \psfrag{arcsin psi2}{{\large $\arcsin |\psi_2|$}}
    \psfrag{arcsin psi2 - theta_m}{{\large $\arcsin|\psi_2|-\theta_m$}}
    \psfrag{(A)}{{\large $(A)$}}
    \psfrag{(B)}{{\large $(B)$}}
    \psfrag{0.9}{{\large $0.9$}}
    \psfrag{1}{{\large $1$}}
    \psfrag{1.1}{{\large $1.1$}}
    \psfrag{1.2}{{\large $1.2$}}
    \psfrag{1.3}{{\large $1.3$}}
    \psfrag{1.4}{{\large $1.4$}}
    \psfrag{1.5}{{\large $1.5$}}
    \psfrag{0}{{\large $0$}}
    \psfrag{0.02}{{\large $0.02$}}
    \psfrag{0.04}{{\large $0.04$}}
    \psfrag{0.06}{{\large $0.06$}}
    \psfrag{0.08}{{\large $0.08$}}
    \psfrag{0.1}{{\large $0.1$}}
    \psfrag{0.12}{{\large $0.12$}}
    \psfrag{0.8}{{\large $0.8$}}
    \psfrag{1}{{\large $1.0$}}
    \psfrag{1.2}{{\large $1.2$}}
    \psfrag{1.4}{{\large $1.4$}}
    \psfrag{Exponential}{\textsf{\large Exponential}}
    \psfrag{BP2000}{\textsf{\large BP2000}}
    \includegraphics[angle=0,width=.98\textwidth]{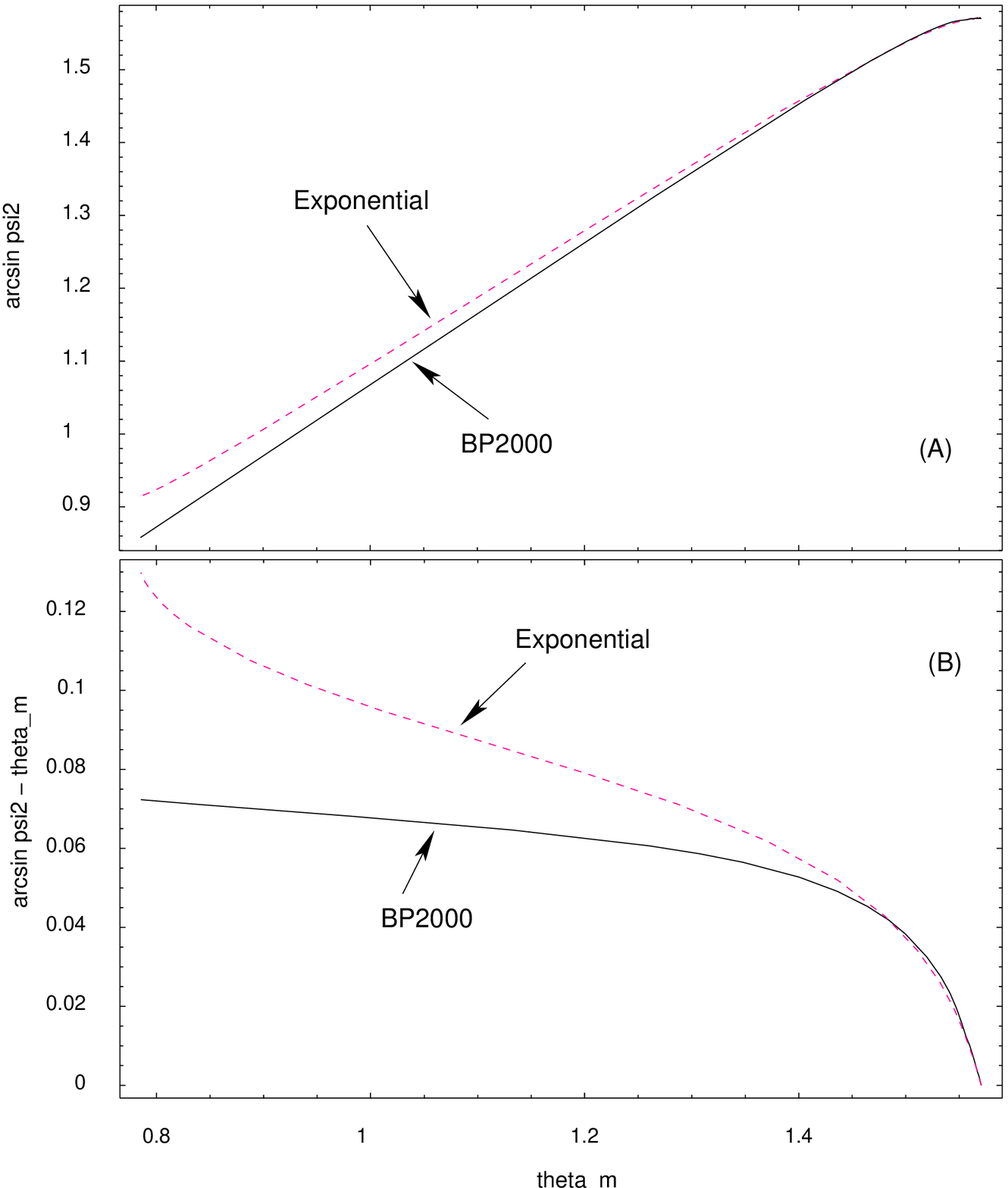}
    \caption{The comparison of the neutrino state evolution in the
      infinite exponential profile (dashed) and the realistic 
      solar profile according to the BP2000 solar model (solid),
      plotted as a function of the mixing angle in matter $\theta_m$. The
      bottom figure shows the deviation of the evolution in both cases
      from extreme nonadiabatic.}
    \label{fig:bp2000vsexp_theta_m}
  \end{center}
\end{figure}

\begin{figure}[htbp]
  \begin{center}
    \psfrag{x, km}{{\large $x$, $\times 10^6$ km}}
    \psfrag{arcsin psi2}{{\large $\arcsin |\psi_2|$}}
    \psfrag{arcsin psi2 - theta_m}{{\large $\arcsin|\psi_2|-\theta_m$}}
    \psfrag{(A)}{{\large $(A)$}}
    \psfrag{(B)}{{\large $(B)$}}
    \psfrag{0.9}{{\large $0.9$}}
    \psfrag{1}{{\large $1$}}
    \psfrag{1.1}{{\large $1.1$}}
    \psfrag{1.2}{{\large $1.2$}}
    \psfrag{1.3}{{\large $1.3$}}
    \psfrag{1.4}{{\large $1.4$}}
    \psfrag{1.5}{{\large $1.5$}}
    \psfrag{0}{{\large $0$}}
    \psfrag{0.02}{{\large $0.02$}}
    \psfrag{0.04}{{\large $0.04$}}
    \psfrag{0.06}{{\large $0.06$}}
    \psfrag{0.08}{{\large $0.08$}}
    \psfrag{0.1}{{\large $0.1$}}
    \psfrag{0.12}{{\large $0.12$}}
    \psfrag{0.2E6}{{\large $0.2$}}
    \psfrag{0.4E6}{{\large $0.4$}}
    \psfrag{0.6E6}{{\large $0.6$}}
    \psfrag{0.8E6}{{\large $0.8$}}
    \psfrag{1E6}{{\large $1.0$}}
    \psfrag{1.2E6}{{\large $1.2$}}
    \psfrag{1.4E6}{{\large $1.4$}}
    \psfrag{Exponential}{\textsf{\large Exponential}}
    \psfrag{BP2000}{\textsf{\large BP2000}}
    \includegraphics[angle=0,width=.98\textwidth]{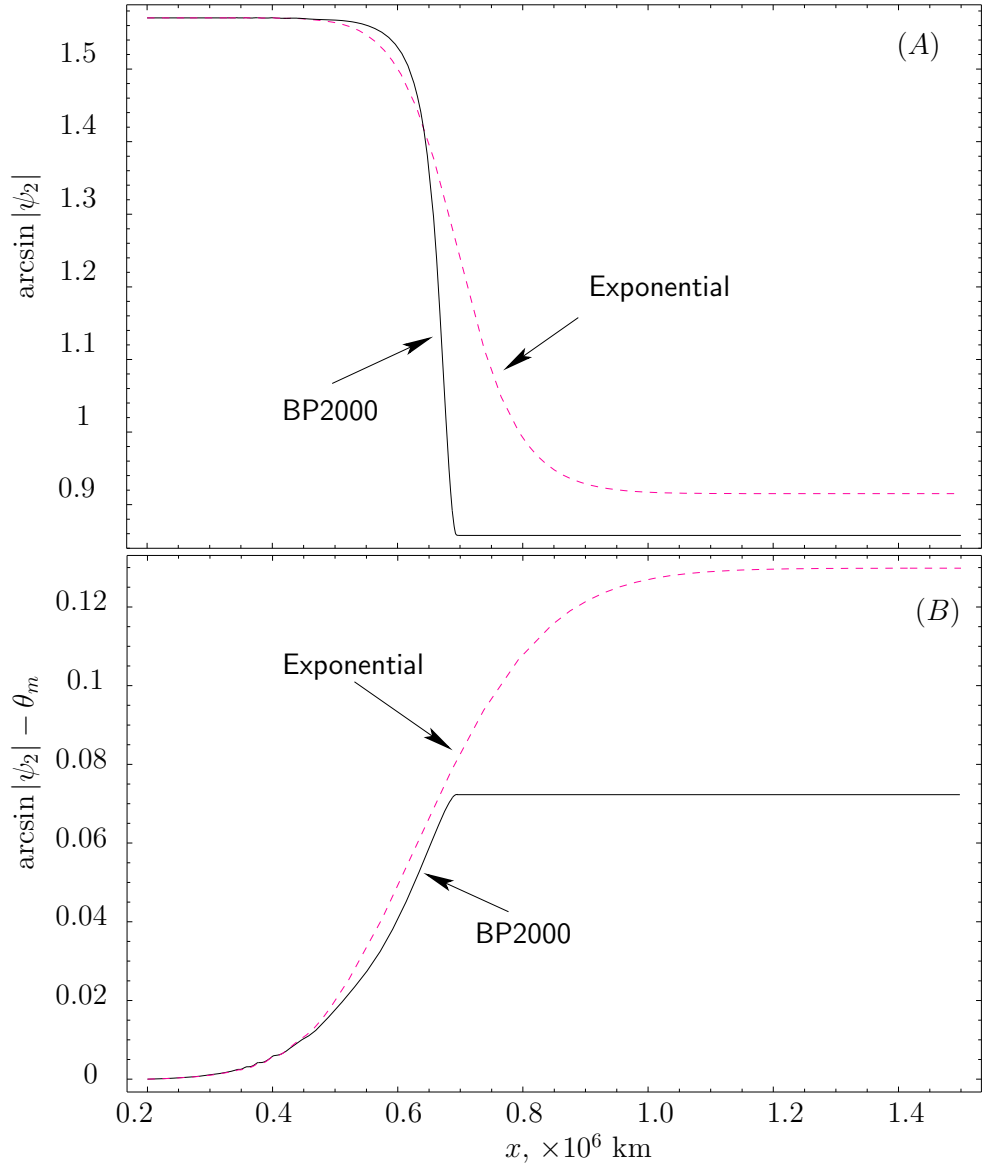}
    \caption{Same as Fig. \ref{fig:bp2000vsexp_theta_m}, but as a
      function of the distance from the center of the Sun $x$.}
    \label{fig:bp2000vsexp_x}
  \end{center}
\end{figure}


In this subsection we discuss the matter effects in the QVO
regime. This region, characterized by $\Delta m^2/E_\nu \sim 10^{-9}$
eV$^2$/MeV and large values of the mixing angle $\theta$ deserves a
separate treatment. While lying below the adiabatic region as
determined by the criterion of Eq. (\ref{eq:adiabaticity}) (see
Fig. \ref{fig:empiricalfit}), it is nonetheless characterized by a
significant deviation from extreme nonadiabaticity. In fact, as
discussed in the last section, for $\Delta m^2/E_\nu \lesssim 5\times
10^{-9}$ eV$^2$/MeV the jumping probability in this region can be
described quite well by Eq. (\ref{eq:expPc}) with
$r_0=R_{\odot}/18.4$. Thus, the neutrino evolution in this region is
\emph{partially adiabatic}, and our task next is to sketch a simple
physical picture of this phenomenon.

First, notice that the condition $A\simeq\Delta$ in the QVO region is
satisfied very close to the Sun's edge (at $x=6.6 \times 10^5$ km for
$\Delta m^2/E_\nu = 10^{-9}$ eV$^2$/MeV), where the profile falls off
very rapidly. This of course does not mean that the evolution is
extremely nonadiabatic. As discussed previously, the slope of the
profile at the point $A=\Delta$ can only be used to find the jumping
probability in the case of small mixing angles. For large mixing
angles, the nonadiabatic segment of the neutrino trajectory is large,
and roughly the first half of this segment lies within 
$x\lesssim 0.92 R_\odot=6.35 \times 10^5$ km, where the density
is equal to or greater than that of the fitted exponential profile.

This suggests that the evolution of the solar neutrino in the QVO
regime can be modeled as the evolution in the exponential profile
truncated near the point of the maximal violation of adiabaticity.
Let us investigate the basic features of this model.

First, consider a partially adiabatic evolution ($\cos\theta
|\Delta_m/\dot\theta_m|_{\theta_m\sim\pi/4+\theta/2} \sim 1$) in case
of the infinite exponential profile. As the neutrino traverses the Sun,
its state vector, although not completely ``attached'' to the heavy
mass eigenstate, rotates away from the pure $\nu_e$ flavor state. This
rotation occurs both before and after the point of the maximal
violation of adiabaticity. Furthermore, an important role is played by
the parts of the trajectory where $\theta_m$ is close to either
$\pi/2$ or $\theta$, where the quantity $\Delta_m/\dot\theta_m$ is
large (see Fig.~\ref{fig:Dmthetadot}).

Now truncate the profile close to the point of the maximal violation of
adiabaticity. Then the amount of the rotation of the state with
respect to the flavor basis decreases roughly by a factor of
two. Actually, because the function in Eq. (\ref{eq:exponential}) has
a \emph{double} pole at $\theta_m=\pi/2$, and a \emph{single} pole
around $\theta_m=\theta$ the factor can be expected to be somewhat
less than two. The numerical calculations of the last section indeed
show that for the BP2000 profile the parameter $r_0$ changes from
$R_\odot/10.54$ to $R_\odot/18.4$.

To illustrate this more quantitatively, let us see how
$|\psi_2(\theta_m)|$ deviates from its extreme nonadiabatic behavior
$|\psi_2(\theta_m)|=\sin\theta_m$ in the QVO regime.
Fig. \ref{fig:bp2000vsexp_theta_m}(A) shows the probability of finding
the neutrino in the heavy mass eigenstate, for $\Delta
m^2/E_\nu=10^{-9}$ eV$^2$/MeV, $\theta=\pi/4$, as a function of the
mixing angle in matter $\theta_m$. Results for both the infinite
exponential profile and the BP2000 profile are shown.
Fig. \ref{fig:bp2000vsexp_theta_m}(B) shows the graph of
$\Delta\theta_m\equiv\arcsin|\psi_2(\theta_m)|-\theta_m$ as a function
of $\theta_m$. The main contribution to $\Delta\theta_m$ comes from
the region $1.4 \lesssim \theta_m <\pi/2$, where the profile is close
to the exponential.

Fig. \ref{fig:bp2000vsexp_x} shows the same evolution in the physical
$x$--space. One can see that most of the contribution to
$\Delta\theta_m$ comes from the part of the profile between
$0.7R_\odot$ and $R_\odot$. This confirms that it would be incorrect
to try to estimate $\Delta\theta_m$ by computing the slope around the
point $A=\Delta$, since for large $\theta$ the entire region $x\gtrsim
0.7 R_\odot$ is important.

\setcounter{equation}{0}
\setcounter{footnote}{0}
\section{All known solutions can be derived from the one for the
  $\tanh$ profile.} 
\label{sect:onesolution}


The formulation of the evolution equations introduced in Section
\ref{sect:intro}, with $\theta_m$ as an independent variable, makes it
possible to uncover a simple relationship between the known analytical
solutions. Such solutions have been obtained for a very limited set of
profiles. In addition to the exponential density distribution,
explicit formulas in the literature have only been given for (i) a
linear density distribution $A(x)=-C_0 x$
\cite{Haxton1986,Parke1986}\footnote{Notice that
Eq. (\ref{eq:linearPc}) is often written in a form containing a factor
of $1/\cos 2\theta$. This is done to express $C_0^{-1}$ in terms of
the logarithmic derivative of the density at the resonance. In that
form, it may superficially appear that $P_c$ has a singularity at
$\theta=\pi/4$. Of course, once the derivative is computed, the
factors of $\cos 2\theta$ cancel out. As we have discussed in this
paper, the conventional definition of the resonance has little
physical meaning for large angles, so there is no good reason for
introducing the factor of $1/\cos 2\theta$.}  ,
\begin{eqnarray}
  \label{eq:linearPc}
  (P_c)_{\rm (lin)}=\exp(-\pi \Delta^2(C_0)^{-1} \sin^2 2\theta),
\end{eqnarray}
(ii) a distribution $A(x)=B_0/x$  \cite{KP1989_1overr}, 
\begin{eqnarray}
  \label{eq:1overxPc}
  (P_c)_{(1/x)}=\frac{\exp(4\pi B_0 \cos^2\theta)-1}{\exp(4\pi B_0)-1},
\end{eqnarray}
and (iii) the hyperbolic tangent distribution \cite{Notzold_tanh} 
$A(x)=A_0[1-\tanh(x/l)]/2$,
\begin{eqnarray}
  \label{eq:tanhPc}
  (P_c)_{(\tanh)}=\frac{\cosh(\pi l A_0)-\cosh(\pi l
    (\Delta_\infty-\Delta))}
  {\cosh(\pi l (\Delta_\infty+\Delta))-\cosh(\pi l (\Delta_\infty-\Delta))}.
\end{eqnarray}
In the last equation $\Delta_\infty$ is the value of $\Delta_m$ deep
inside matter, $ \Delta_\infty\equiv \lim_{x\rightarrow\infty}
\Delta_m=\sqrt{A_0^2-2 A_0 \Delta\cos 2\theta+\Delta^2}$.

What do these distributions have in common that makes them exactly solvable?
One important feature that unites them is that the
corresponding differential equations can all be put in the
hypergeometric form \cite{BychukSpiridonov}.
We next show more directly that Eqs. (\ref{eq:expPc}), (\ref{eq:linearPc}),
(\ref{eq:1overxPc}), and (\ref{eq:tanhPc}) are all related to each
other and that the first three can be easily obtained from the last one. 

To make this relationship clear, it is useful to show
the analogs of Eq. (\ref{eq:exponential}) for the other three
distributions in question. A straightforward calculation yields
\begin{eqnarray}
  \label{eq:linearDmthetadot}
  \left(\frac{\Delta_m}{\dot\theta_m}\right)_{\rm (lin)} = 
  -\frac{2\Delta^2 \sin^2 2\theta}{C_0\sin^3 2\theta_m}
\end{eqnarray}
for the linear distribution,
\begin{eqnarray}
  \label{eq:1overxDmthetadot}
  \left(\frac{\Delta_m}{\dot\theta_m}\right)_{(1/x)} = 
  -\frac{2 B_0 \sin^2 2\theta}{\sin 2\theta_m \sin^2 (2\theta_m-2\theta)}
\end{eqnarray}
for the $1/x$ distribution, and
\begin{eqnarray}
  \label{eq:tanhDmthetadot}
  \left(\frac{\Delta_m}{\dot\theta_m}\right)_{(\tanh)} = 
  \frac{A_0 l \sin 2\theta\sin 2\theta_\infty}
  {\sin 2\theta_m \sin(2\theta_m-2\theta)\sin(2\theta_m-2\theta_\infty)}
\end{eqnarray}
for the hyperbolic tangent distribution ($\theta_\infty$ is the value
of $\theta_m$ deep inside matter, so that
$\sin 2\theta_\infty=\Delta \sin 2\theta/\Delta_\infty$).

By inspecting Eqs. (\ref{eq:exponential}),
(\ref{eq:linearDmthetadot}), (\ref{eq:1overxDmthetadot}), and
(\ref{eq:tanhDmthetadot}) one can clearly see the common origin of all
four solutions. For the hyperbolic tangent distribution
$(\Delta_m/\dot{\theta}_m)^{(\tanh)}$ has three simple poles in
$\theta_m$ on $(0,\pi/2]$, thus representing the most general case of
all four. The exponential case is obtained when two of the poles,
$\theta_\infty$ and $\pi/2$, merge in one double pole, and the $1/x$
case is obtained when the poles $\theta$ and $\theta_\infty$
merge. Finally, if all three poles merge in one triple pole at $0$
($\pi/2$), one obtains the linear case. Thus, given the result for the
hyperbolic tangent distribution, it should be possible to recover the
answers for the other three distributions by simply taking appropriate
limits.

We first show how the
exponential result can be obtained from the one for the 
hyperbolic tangent. By comparing Eqs. (\ref{eq:expPc}) and
(\ref{eq:tanhPc}) we see that we need a large $A_0$ limit of
(\ref{eq:tanhPc}), since in this case\footnote{Notice that the product
$A_0 \sin 2\theta_\infty$ in the limit of large $A_0$ approaches a
constant value $\Delta\sin 2\theta$.} $\theta_\infty\rightarrow\pi/2$,
and also a substitution $l\rightarrow 2r_0$.
\begin{eqnarray}
  \label{eq:expfromtanh}
  (P_c)_{(\tanh)}&\longrightarrow& \frac{\exp(\pi l A_0)/2-\exp(\pi l
    (\Delta_\infty-\Delta))/2}
  {\exp(\pi l (\Delta_\infty+\Delta))/2-\exp(\pi l
    (\Delta_\infty-\Delta))/2}\nonumber\\
  &=&\frac{\exp(\pi l (A_0-(\Delta_\infty-\Delta)))-1}
  {\exp(\pi l (2\Delta))-1}\nonumber\\
  &\longrightarrow&
  \frac{\exp(\pi l(A_0-A_0(1-(\Delta/A_0)\cos 2\theta)+\Delta))-1}
  {\exp(\pi l (2\Delta))-1}\nonumber\\
  &=&\frac{\exp(\pi l\Delta(1+\cos 2\theta))-1}
  {\exp(\pi l (2\Delta))-1}\nonumber\\
  &\longrightarrow& \frac{\exp(4\pi r_0\Delta\cos^2\theta)-1}
  {\exp(4\pi r_0 \Delta)-1}=(P_c)_{\rm exp}
\end{eqnarray}
Above we used the fact that for large $A_0$ $\Delta_m=\sqrt{A_0^2-2
A_0 \Delta\cos 2\theta+\Delta^2}\rightarrow A_0 (1-(\Delta/A_0)\cos
2\theta)$.  The physical interpretation of this result is the
following. For a fixed $\Delta$ and $A_0\rightarrow \infty$ the part
of the neutrino trajectory where adiabaticity is maximally
violated occurs at large $x$, where
$A_0[1+\tanh(x/l)]/2\rightarrow A_0 \exp(-2x/l)=A_0 \exp(-x/r_0)$.
Notice, that $A_0$ dropped out and the derivation is valid for all
$\Delta$ and $\theta$.

To obtain the expression for $P_c$ for the linear profile from that
for the exponential, in Eq. (\ref{eq:exponential}) we take the limit
$r_0\rightarrow\infty$, $\theta\rightarrow 0$, such that the product
$2\Delta r_0 \sin^2 2\theta$ approaches a constant value,
$2(\tilde{\Delta}^2/C_0)\sin^2 2\tilde{\theta}$. In this limit
$\exp(4\pi r_0\Delta \cos^2 \theta)\gg 1$ and $\sin^2
2\theta\rightarrow 4\sin^2\theta$, so that
\begin{eqnarray}
  \label{eq:linfromexp}
  (P_c)_{\rm exp}&=&\frac{\exp(4\pi r_0\Delta\cos^2\theta)-1}
  {\exp(4\pi r_0\Delta)-1}
  \longrightarrow\exp(-4\pi r_0\Delta\sin^2\theta)\nonumber\\
  &\longrightarrow&\exp(-\pi (\tilde{\Delta}^2/C_0)\sin^2
2\tilde{\theta})=(P_c)_{\rm lin}.
\end{eqnarray}

At last, we will show how the result for the $1/x$ distribution
follows from that for the hyperbolic tangent distribution. The logic
is similar to what was done before: Eq. (\ref{eq:1overxDmthetadot})
can be obtained from Eq. (\ref{eq:tanhDmthetadot}) by sending
$\theta_\infty\rightarrow\theta_m$ and relabeling $A_0 l \rightarrow
2A_0$; the result for the $1/x$ profile should then be read off from
the solution of the differential equation with the profile
(\ref{eq:tanhDmthetadot}). A complication arises because in this case
one needs to know the solution of the differential equation
\emph{between} $\pi/2$ and $\theta_\infty$, while the known result,
Eq. (\ref{eq:tanhPc}) describes the solution between $\theta_\infty$
and $\theta$. The range $[\theta_\infty,\pi/2]$ (and also
$[0,\theta]$) correspond to a different matter distribution,
$A(x)=A_0[1+1/\tanh(x/l)]/2$ (see Appendix \ref{sect:appendix}).

This difficulty can be easily resolved and the result in question can
be obtained from Eq. (\ref{eq:tanhPc}) by appropriate
substitutions. The key observation is that in the case of the
hyperbolic tangent the differential equation is uniquely specified by
the \emph{relative} positions of the three poles of
Eq. (\ref{eq:tanhDmthetadot}) and the factor in the numerator. By
shifting the poles such that $\theta\rightarrow 0$, we can reduce the
problem of finding the solution on $[\theta_\infty,\pi/2]$ to the
solved case $[\theta',\theta'_\infty]$ and use Eq. (\ref{eq:tanhPc}).

The details of this procedure can be found in Appendix
\ref{sect:appendix}. After a straightforward calculation one finds that
\begin{eqnarray}
  \label{eq:Pcnewprofile}
\lefteqn{P'_c=}\nonumber\\
&&\frac{
  \cosh\left(\pi l A_0\frac{\sin 2\theta_\infty}{\sin
    (2\theta_\infty-2\theta)}\right)-
   \cosh\left(\pi l A_0\left(\frac{\sin 2\theta}{\sin
    (2\theta_\infty-2\theta)}-1\right)\right)}
  {\cosh\left(\pi l A_0\left(\frac{\sin 2\theta}{\sin
    (2\theta_\infty-2\theta)}+1\right)\right)-
   \cosh\left(\pi l A_0\left(\frac{\sin 2\theta}{\sin
    (2\theta_\infty-2\theta)}-1\right)\right)}=\nonumber\\
&&\frac{
  \cosh\left(\pi l \Delta\right)-
   \cosh\left(\pi l (\Delta_\infty-A_0)\right)}
  {\cosh\left(\pi l (\Delta_\infty+A_0)\right)-
   \cosh\left(\pi l (\Delta_\infty-A_0)\right)}
\end{eqnarray}
From the first form of the expression it is easy to see that the
formula satisfies the necessary nonadiabatic limit: $\lim_{l
\rightarrow 0} P'_c =\cos^2\theta$. From the second form it is easy to
take the limit which reproduces the formula for the $1/x$ profile. We
can achieve $\theta_\infty\rightarrow\theta$ by making $\Delta$ large
while keeping $A_0$ fixed. Once again, in this limit
$\Delta_\infty\rightarrow\Delta-A\cos 2\theta$ and so
\begin{eqnarray}
  \label{eq:1overxfromtanh}
  P'_c&\longrightarrow& \frac{
  \exp(\pi l \Delta)-
   \exp(\pi l (\Delta_\infty-A_0))}
  {\exp(\pi l (\Delta_\infty+A_0))-
   \exp(\pi l (\Delta_\infty-A_0))}\nonumber\\
 &\longrightarrow&
\frac{\exp(\pi l (\Delta-\Delta_\infty+A_0))-1}
{\exp(2\pi l A_0)-1}\nonumber\\
&\longrightarrow& 
\frac{\exp(2\pi l A_0 \cos^2\theta)-1}{\exp(2\pi l A_0)-1}.
\end{eqnarray}
Upon relabeling $A_0 l \rightarrow 2B_0$ we recover
Eq. (\ref{eq:1overxPc}).  

The last result deserves a few comments. It is quite remarkable that
the dependence of the jumping probability on the mass-squared
splitting and energy completely dropped out at the end. The $1/x$
profile thus represents a rather unique case when the adiabaticity is
entirely determined by the density profile and the mixing angle. This
could have been, of course, anticipated already on the basis of the
dimensional analysis.  Indeed, the parameter $B_0$ is
\emph{dimensionless} and thus, together with $\theta$, completely
determines the jumping probability.

It is instructive to see how the cancellation happens for small
$\theta$ and $B_0\gg 1$.  Eq. (\ref{eq:1overxPc}) then becomes
$P_c=\exp(-4\pi B_0\theta^2)$.  But we can also compute this
differently: since for small angles the resonance is narrow, we can
use a linear formula 
$P_c= \exp(-\pi \Delta^2|dA(x)/dx|_{\rm res}^{-1}\sin^2 2\theta)$. 
Since for $A(x)=B_0/x$ we have $dA(x)/dx=-A(x)^2/B_0$ and at the point
of resonance $A(x)=\Delta$, the quantity $\Delta$ cancels out of the
final result: 
\begin{equation}
  P_c=\exp(-\pi \Delta^2 4\theta^2 B_0/\Delta^2)
  =\exp(-4\pi \theta^2 B_0)
\end{equation}
The slope at the resonance point changes exactly in such a way as to
compensate for the change in $\Delta^2$. The solar neutrino fits would
be qualitatively different, were the solar density profile close to
$1/x$ instead of the exponential.

\setcounter{footnote}{0}
\section{Conclusions}


In this paper we have discussed several aspects of the neutrino evolution
in matter, emphasizing the case of the large values of
the mixing angle, including values greater than $\pi/4$. We have
pointed out how some of the results originally derived for small
mixing angles can be modified to be applicable to all values of
$\theta$. Such results include the adiabaticity condition and the role
played by the resonance in determining where the nonadiabatic jumping
between the states takes place. We have formulated analytical criteria
for the case of the exponential matter distribution and commented on
how these criteria apply to case of the realistic solar profile.
Although the focus of our analysis was on solar neutrinos, the results
are useful for understanding the physics of neutrino oscillations in
matter in general.

We have presented the results of accurate numerical calculations,
showing how the jumping probability $P_c$ interpolates between the QVO
and the standard MSW regimes in the case of the realistic solar
profile. An empirical prescription on how to estimate $P_c$ anywhere
in this range with only elementary functions was given. The matter
effects in the quasivacuum regime were discussed.

Finally, we have shown that the known analytical solutions for the
linear, exponential, and $1/x$ density distributions can be
easily obtained from the result for the hyperbolic tangent. It was
especially easy to see this using $\theta_m$ as an independent
variable. In the process of the proof we also obtained an answer for
a fifth distribution, $N_e \propto (\coth(x/l)\pm 1)$.

\section*{Acknowledgments}

I would like to thank Plamen Krastev for many pleasant and
enlightening conversations and for pointing out several very important
references. I am also grateful to Hitoshi Murayama and John Bahcall
for their support.  
This work was supported by the Keck Foundation and by the Director,
Office of Science, Office of High Energy and Nuclear Physics, Division
of High Energy Physics of the U.S. Department of Energy under
Contracts DE-AC03-76SF00098. 

\appendix
\setcounter{equation}{0}
\section{Derivation of the expression for $P_c$ for the 
   density distribution $A(x)\propto(\coth(x/l)\pm 1)$}
\label{sect:appendix}

As mentioned in Section \ref{sect:onesolution}, one of the four matter
distributions for which exact expressions for $P_c$ have been obtained
is the profile $A(x)=A_0 [1+\tanh(x/l)]/2$. The answer is given by
Eq. (\ref{eq:tanhPc}) and represents the result of solving the
differential equation (\ref{eq:thetavariable}) on the interval
$[\theta,\theta_\infty]$, with $\Delta_m/\dot\theta_m$ given by
Eq. (\ref{eq:tanhDmthetadot}).  In this Appendix we show how this
result can be used to obtain the solution to the same differential
equation on the interval $[\theta_\infty,\pi/2]$.

Clearly for any $x$ in the distribution $A(x)=A_0 [1+\tanh(x/l)]/2$
the angle $\theta_m$ is constrained between $\theta$ and
$\theta_\infty$.  The range of $\theta_m$ $[\theta_\infty,\pi/2]$
corresponds to a different density profile, which everywhere satisfies 
$A(x)>A_0$. This new profile can be found by simple analytical
continuation. In the original profile the value of the density
$A(x)=C$ occurs at
\begin{eqnarray}
  \label{eq:arctanh}
  x=\frac{l}{2}\log\frac{C}{A_0-C}.
\end{eqnarray}
When $C>A_0$ the argument of the logarithm becomes negative. Using
the analytical continuation of the logarithm, it can be interpreted as
\begin{eqnarray}
  \label{eq:arctanh2}
  x=\frac{l}{2}\left(\log\frac{C}{|A_0-C|}+i\pi\right)=
  \tilde{x}+\frac{i\pi l}{2}.
\end{eqnarray}
Substituting this in the equation for $A(x)$ we find
\begin{eqnarray}
  \label{eq:Atilde}
  \tilde{A}(\tilde{x})=A_0[1+\tanh(\tilde{x}/l+i\pi/2)]/2=
  A_0[1+1/\tanh(\tilde{x}/l)]/2.
\end{eqnarray}

To obtain the expression for $P_c$ for the distribution
(\ref{eq:Atilde}), as a first step in Eq. (\ref{eq:tanhPc}) we express
the quantities $\Delta_\infty$ and $\Delta_m$ in terms of the angles
$\theta$ and $\theta_m$,
\begin{eqnarray}
  \label{eq:tahnPc2}
    (P_c)_{(\tanh)}=
    \frac{\cosh(\pi l A_0)-
      \cosh\left(\pi l A_0\frac{\sin 2\theta_\infty-\sin 2\theta}
        {\sin(2\theta_\infty-2\theta)}\right)}
    {\cosh\left(\pi l A_0\frac{\sin 2\theta_\infty+\sin 2\theta}
      {\sin(2\theta_\infty-2\theta)}\right)-
  \cosh\left(\pi l A_0\frac{\sin 2\theta_\infty-\sin 2\theta}
    {\sin(2\theta_\infty-2\theta)}\right)}.
\end{eqnarray}

Next we shift $\theta_m\rightarrow\theta_m-\theta$ in
Eq. (\ref{eq:tanhDmthetadot}) such that the pole at $\theta$ moves
to $0$. This reduces the problem to solving the differential
equation (\ref{eq:tanhDmthetadot}) between
$\theta'_\infty\equiv\pi/2-\theta$ and
$\theta'\equiv\theta_\infty-\theta$. 
To complete the change to the primed variables, in the numerator of
Eq. (\ref{eq:tanhDmthetadot}) we substitute $A_0 \sin 2\theta\sin
2\theta_\infty$ by $A'_0 \sin 2\theta'\sin 2\theta'_\infty$, where
$A'_0=A_0 \sin 2\theta_\infty/\sin (2\theta_\infty-2\theta)$. 
We then find
\begin{eqnarray}
  \label{eq:Pcprm}
  \lefteqn{P'_c=}\nonumber\\
    &&\frac{\cosh(\pi l A'_0)-
      \cosh\left(\pi l A'_0\frac{\sin 2\theta'_\infty-\sin 2\theta'}
        {\sin(2\theta'_\infty-2\theta')}\right)}
    {\cosh\left(\pi l A'_0\frac{\sin 2\theta'_\infty+\sin 2\theta'}
      {\sin(2\theta'_\infty-2\theta')}\right)-
  \cosh\left(\pi l A'_0\frac{\sin 2\theta'_\infty-\sin 2\theta'}
    {\sin(2\theta'_\infty-2\theta')}\right)}=\nonumber\\
&& \frac{\cosh\left(\pi l A_0\frac{\sin 2\theta_\infty}
  {\sin (2\theta_\infty-2\theta)}\right)-
      \cosh\left(\pi l A_0\left(\frac{\sin 2\theta}
  {\sin (2\theta_\infty-2\theta)}-1\right)\right)}
    {\cosh\left(\pi l A_0\left(\frac{\sin 2\theta}
  {\sin (2\theta_\infty-2\theta)}+1\right)\right)-
  \cosh\left(\pi l A_0\left(\frac{\sin 2\theta}
  {\sin (2\theta_\infty-2\theta)}-1\right)\right)}.\;\;\;\;\;\;
\end{eqnarray}
This is the answer that is needed in Sect. \ref{sect:onesolution}. 

One may be interested in the expression for $P_c$ for a
distribution $A(x)=A_0[1/\tanh(x/l)-1]/2$, which has the property that 
it vanishes in the limit of large $x$. This can be found by redefining
the vacuum values of the neutrino oscillation parameters to be
$\Delta_\infty$ and $\theta_\infty$. Eq. (\ref{eq:Pcprm}) can then be
rewritten as
\begin{eqnarray}
  \label{eq:Pcprm2}
  P'_c=\frac{
  \cosh\left(\pi l \Delta\right)-
   \cosh\left(\pi l (\Delta_\infty-A_0)\right)}
  {\cosh\left(\pi l (\Delta_\infty+A_0)\right)-
   \cosh\left(\pi l (\Delta_\infty-A_0)\right)},
\end{eqnarray}
where $\Delta=\sqrt{A_0^2+2 A_0 \Delta_\infty\cos
  2\theta_\infty+\Delta_\infty^2}$.

\end{document}